\documentclass[conference]{IEEEtran}
\IEEEoverridecommandlockouts
\usepackage{cite}
\usepackage{amsmath,amssymb,amsfonts}
\usepackage{algorithmic}
\usepackage{graphicx}
\usepackage{textcomp}
\usepackage{xcolor}
\usepackage{hyperref}
\def\BibTeX{{\rm B\kern-.05em{\sc i\kern-.025em b}\kern-.08em
    T\kern-.1667em\lower.7ex\hbox{E}\kern-.125emX}}

\usepackage{amsmath,amssymb,amsfonts}
\usepackage{graphicx}
\usepackage{textcomp}
\usepackage{xcolor}
\usepackage{enumitem}
\usepackage{xurl}
\usepackage{balance}
\usepackage{lastpage}

\usepackage{lipsum,multicol}
\usepackage[many]{tcolorbox}
\usepackage{wrapfig}
\usepackage{url}
\usepackage{float}
\usepackage{listings}
\usepackage[labelsep=colon,font=footnotesize]{caption}
\usepackage{xspace}
\usepackage{caption}
\usepackage{subcaption}
\usepackage{tikz}
\usepackage{scalerel}
\usetikzlibrary{svg.path}
\usepackage{multirow}
\usepackage{colortbl}
\usepackage{booktabs}

\definecolor{orcidlogocol}{HTML}{A6CE39}
\tikzset{
  orcidlogo/.pic={
    \fill[orcidlogocol] svg{M256,128c0,70.7-57.3,128-128,128C57.3,256,0,198.7,0,128C0,57.3,57.3,0,128,0C198.7,0,256,57.3,256,128z};
    \fill[white] svg{M86.3,186.2H70.9V79.1h15.4v48.4V186.2z}
                 svg{M108.9,79.1h41.6c39.6,0,57,28.3,57,53.6c0,27.5-21.5,53.6-56.8,53.6h-41.8V79.1z M124.3,172.4h24.5c34.9,0,42.9-26.5,42.9-39.7c0-21.5-13.7-39.7-43.7-39.7h-23.7V172.4z}
                 svg{M88.7,56.8c0,5.5-4.5,10.1-10.1,10.1c-5.6,0-10.1-4.6-10.1-10.1c0-5.6,4.5-10.1,10.1-10.1C84.2,46.7,88.7,51.3,88.7,56.8z};
  }
}

\newcommand\orcidicon[1]{\href{https://orcid.org/#1}{\mbox{\scalerel*{
\begin{tikzpicture}[yscale=-1,transform shape]
\pic{orcidlogo};
\end{tikzpicture}
}{|}}}}

\usepackage{hyperref} 

\newtcolorbox{boxB}{
    fontupper = \bf\color{main}\footnotesize, 
    boxrule = 0.5pt,
    colframe = main,
    rounded corners,
    arc = 5pt   
}

\newtcolorbox{boxD}{
    fontupper = \small, 
    colback = sub, 
    colframe = main, 
    boxrule = 0pt, 
    toprule = 2pt, 
    bottomrule = 2pt 
}

\newtcolorbox{boxH}{
    fontupper = \small, 
    colback = sub, 
    colframe = main, 
    boxrule = 0pt, 
    leftrule = 6pt 
}

\newtcolorbox{boxG}{
    enhanced,
    boxrule = 0pt,
    colback = sub,
    borderline west = {1pt}{0pt}{main}, 
    borderline west = {0.75pt}{2pt}{main}, 
    borderline east = {1pt}{0pt}{main}, 
    borderline east = {0.75pt}{2pt}{main}
}    

\newtcolorbox{boxK}{
    fontupper = \small,
    sharpish corners, 
    boxrule = 0pt,
    toprule = 1.0pt, 
    enhanced,
    fuzzy shadow = {0pt}{-2pt}{-0.5pt}{0.5pt}{black!35} 
}

\newcommand*\circled[1]{\tikz[baseline=(char.base)]{
            \node[shape=circle,draw,inner sep=0.5pt] (char) {#1};}}

\newboolean{changes}

\setboolean{changes}{true}
\ifthenelse{\boolean{changes}}
{
}

\newcommand{\ie}{\textit{i.e.,}\xspace}
\newcommand{\eg}{\textit{e.g.,}\xspace}

\newcommand{\etal}{et al.\xspace}

\usepackage{pifont}

\newcounter{rcounter}

\newcounter{fcounter}

\newcommand{\secref}[1]{Sec.~\ref{#1}\xspace}
\newcommand{\figref}[1]{Fig.~\ref{#1}\xspace}
\newcommand{\tabref}[1]{Tab.~\ref{#1}\xspace}

\newtcolorbox{story}[1][]{
  width=\textwidth,
  fonttitle=\bfseries,
  breakable,
  fonttitle=\bfseries\color{Brown},
  colframe=Melon,
  colback=Melon!10
  #1}

\newtcolorbox[use counter=mynote]
  {mynote}[1][]
  {title=Note~\thetcbcounter,
   width=0.45\textwidth,
   left=0pt,
   right=0pt,
   fonttitle=\bfseries,
   coltitle=black,
   colframe=lightgray,
   colback=white,
   #1
}

\newcommand{\llm}{LLM\xspace}
\newcommand{\llms}{LLMs\xspace}
\newcommand{\bigvul}{\textit{Big-Vul}\xspace}
\newcommand{\noncloud}{\textit{open-source}\xspace}
\newcommand{\cisco}{\textit{Propietary}\xspace}

\newcommand{\codellama}{\textit{CodeLlama2}\xspace}
\newcommand{\codellamaS}{\textit{CodeLlama2-7B}\xspace}
\newcommand{\codellamaL}{\textit{CodeLlama2-13B}\xspace}
\newcommand{\mistral}{\textit{Mistral-7B}\xspace}

\newcommand{\gptoss}{\textit{gpt-oss}\xspace}
\newcommand{\gptf}{\textit{GPT4o-mini}\xspace}

\definecolor{main}{HTML}{5989cf}    
\definecolor{sub}{HTML}{cde4ff}     

\newcommand{\metricMSE}{\textit{MSE}\xspace}
\newcommand{\metricNonFeasible}{Non-\textit{Feasible}\xspace}
\newcommand{\metricOutofRange}{\textit{Out-Of-Range}\xspace}
\newcommand{\metricExMean}{Ex.\textit{Mean}\xspace}
\newcommand{\metricExStd}{Ex.\textit{Std.}\xspace}
\newcommand{\metricGTMean}{GT.\textit{Mean}\xspace}
\newcommand{\metricGTStd}{GT.\textit{Std.}\xspace}

\newcommand{\short}{\textit{short}\xspace}
\newcommand{\medium}{\textit{medium}\xspace}

\newboolean{revision}
\setboolean{revision}{false}

\ifthenelse{\boolean{revision}}
  {\newcommand\rev[1]{{\textcolor{blue}{#1}}}
  }
  {\newcommand\rev[1]{{{#1}}}
  }
    
\begin{document}
%
\title{
On Predicting Vulnerability Severity Using In-Context Learning: An Industrial Case Study
}
%
%
%
%

%
\makeatletter
\newcommand{\linebreakand}{%
  \end{@IEEEauthorhalign}
  \hfill\mbox{}\par
  \mbox{}\hfill\begin{@IEEEauthorhalign}
}
    \def\balanceissued{unbalanced}
    \let\oldbibitem\bibitem
    \def\bibitem{%
        \ifnum\thepage=\getpagerefnumber{LastPage}%
            \expandafter\ifx\expandafter\relax\balanceissued\relax\else%
                \balance%
                \gdef\balanceissued{\relax}\fi%
            \else\fi%
        \oldbibitem}
\makeatother

\author{
  \IEEEauthorblockN{Daniel Rodriguez-Cardenas\, \orcidicon{0000-0002-3238-1229}\ }
  \IEEEauthorblockA{
    \textit{William \& Mary}\\
    Williamsburg, VA, USA \\
    dhrodriguezcar@wm.edu}
  \and

  \IEEEauthorblockN{David Nader Palacio\, \orcidicon{0000-0001-9683-5616}\ }
  \IEEEauthorblockA{
    \textit{Microsoft}\\
    Seattle, WA, USA \\
    dnader@microsoft.com}
  \and

  \IEEEauthorblockN{Anna Schmedding}
  \IEEEauthorblockA{
    \textit{William \& Mary}\\
    Williamsburg, VA, USA \\
    akschmedding@wm.edu}
  \linebreakand

  \IEEEauthorblockN{Yiyang Lu}
  \IEEEauthorblockA{
    \textit{William \& Mary}\\
    Williamsburg, VA, USA \\
    ylu21@wm.edu}
  \and

  \IEEEauthorblockN{Aadil Mallick}
  \IEEEauthorblockA{
    \textit{William \& Mary}\\
    Williamsburg, VA, USA \\
    amallick@wm.edu}
  \and

  \IEEEauthorblockN{Bill Hudson}
  \IEEEauthorblockA{
    \textit{Cisco Systems}\\
    San Jose, CA, USA \\
    bhudson@cisco.com}
  \linebreakand

  \IEEEauthorblockN{Chris Gourley}
  \IEEEauthorblockA{
    \textit{Cisco Systems}\\
    San Jose, CA, USA \\
    chgourle@cisco.com}
  \and

  \IEEEauthorblockN{Michael Roytman}
  \IEEEauthorblockA{
    \textit{Cisco Systems}\\
    San Jose, CA, USA \\
    roytman@cisco.com}
  \and

  \IEEEauthorblockN{Chris Shenefiel}
  \IEEEauthorblockA{
    \textit{William \& Mary}\\
    Williamsburg, VA, USA \\
    cashenefiel@wm.edu}
  \linebreakand

  \IEEEauthorblockN{Evgenia Smirni}
  \IEEEauthorblockA{
    \textit{William \& Mary}\\
    Williamsburg, VA, USA \\
    esmirni@wm.edu}
  \and

  \IEEEauthorblockN{Denys Poshyvanyk\, \orcidicon{0000-0002-5626-7586}\ }
  \IEEEauthorblockA{
    \textit{William \& Mary}\\
    Williamsburg, VA, USA \\
    denys@cs.wm.edu}
}

\maketitle

\begin{abstract}
Modern software systems require earlier and more scalable vulnerability severity assessment to reduce exposure to high‑impact security flaws. Security analysts typically assign CVSS scores, but this manual triage does not scale with the growth of disclosed vulnerabilities and often depends on cloud LLM services that raise confidentiality concerns. This paper presents an industrial case study on predicting CVSS v3.1 scores directly from vulnerable C/C++ snippets using \textit{in‑context learning} with locally deployable, \noncloud LLMs. We compare \cisco data with the \bigvul dataset, showing sufficiently aligned CVSS distributions to justify \bigvul as a proxy for industrial data when constructing prompt‑based testbeds. We then vary in‑context configurations and model parameters, evaluating \codellamaS, \codellamaL, \mistral, \gptoss, and \gptf using mean squared error (MSE) and feasibility metrics. Our results show that medium‑sized open‑source code models, particularly \codellamaS, can approximate the best cloud performance for CVSS regression when guided by lightweight, output‑constraining prompts, offering a practical, privacy‑preserving building block for severity triage in industrial settings.

\end{abstract}

\begin{IEEEkeywords}
Security, Large Language Models, Vulnerability, Prediction, Severity
\end{IEEEkeywords}


%


\section{Introduction}


Detecting risky flaws in software has become an instrumental practice for large companies and start-ups alike, as modern systems rapidly evolve. \rev{Attackers exploit these vulnerabilities, leading to more complex issues such as data leakage, service interruptions, and increased operational costs~\cite{ibm2024costbreach,ibm2024costbreachdetail}. In industry, vulnerability management pipelines typically involve identifying, prioritizing, and repairing vulnerabilities in various artifacts such as code, configuration files, documentation, and architectural diagrams. Prioritization is based on severity estimation through standardized risk metrics, among which the Common Vulnerability Scoring System (CVSS) has emerged as the de facto severity score ‐, widely adopted by major vendors and the National Vulnerability Database (NVD)~\cite{FIRST_CVSS,CVSS_UserGuide,infosecpros_cvss_2025}.}


\rev{Performing accurate CVSS score prediction \cite{FIRST_CVSS, CVSS_UserGuide} is both critical and challenging. Studies show that only a small fraction of exploited vulnerabilities are initially scored and prioritized correctly, and that roughly half of real‑world exploits occur before CVSS scores are published, creating a dangerous window in which organizations operate with incomplete severity information. At the same time, alternative risk scores, such as Cisco’s Kenna Risk\cite{vmware_kenna_risk_score} and the Exploit Prediction Scoring System (EPSS)~\cite{FIRST_CVSS} build directly on CVSS to quantify exploitability and business impact, confirming the central role of severity in downstream risk models. However, assigning CVSS scores remains a manual and time-consuming task ‐ that does not scale with the growing volume of disclosed vulnerabilities and code changes.}

\rev{Recent advances in Large Language Models (LLMs) suggest that AI systems can assist analysts by classifying vulnerabilities, explaining flaws\cite{li_vulnerability_2021,wu_code_2022,fu_linevul_2022}, and even suggesting code fixes. Cloud ‐ hosted assistants, such as ChatGPT\cite{fu_chatgpt_2023}, Gemini\cite{gemini2023}, and Claude\cite{anthropic2023claude} have demonstrated strong performance on security‑related tasks, but their use in practice is constrained by confidentiality, compliance, and data residency requirements that prevent organizations from uploading proprietary source code and incident data to third ‐ provider \cite{wu_code_2022,CVSS_UserGuide}. These constraints have sparked interest in \noncloud, locally deployable LLMs that can run on customer infrastructure and operate on internal codebases without exposing sensitive artifacts. Although several works have explored machine learning and deep learning models for CVSS prediction using textual descriptions, little is known about how code ‐centric open‑source LLMs perform in severity regression when guided by in‑context learning.}


\rev{This paper investigates whether \textit{in‑context learning} with \noncloud code LLMs can support earlier, privacy‑preserving CVSS v3.1\cite{FIRST_CVSS} severity prediction directly from vulnerable C/C++ snippets. We first compare a \cisco dataset with the public \bigvul corpus and show that their CVSS distributions overlap in the 4–8 range and exhibit similar correlations with adjusted scores, supporting the use of \bigvul as a proxy for industrial data when constructing prompt‑based testbeds. Building on this foundation, we design an in‑context learning pipeline that assembles prompts from vulnerable code, CVSS scores, and descriptions, and systematically vary (i) context configuration ($C_1$–$C_3$), (ii) sequence length, (iii) temperature, and (iv) number of shots. We evaluate multiple \noncloud (\ie \codellamaS, \codellamaL,\mistral,\gptoss) and the cloud model \gptf as a reference using mean squared error (MSE), \metricNonFeasible output rate and \metricOutofRange predictions as our primary metrics. Our results show that \codellamaS is the most reliable open‑source option, if a vulnerability inspector wants a local and small option, achieving \metricMSE around 7 with three‑shot prompts and virtually \metricNonFeasible outputs, while \gptf under a constrained context ($C_3$) attains substantially lower \metricMSE at the cost of requiring cloud deployment.}

\rev{In this work, we restrict ourselves to few‑shot prompting, focusing on simple \textit{in‑context} configurations rather than more sophisticated reasoning pipelines. Although this keeps our setup aligned with practical, low‑latency deployment constraints, it also leaves open the opportunity to explore richer prompting strategies—such as chain‑of‑thought, self‑reflection, or “thinking” models with built‑in reasoning capabilities—as an important direction for future work to further improve the reliability and interpretability of severity predictions.}

In summary, this work makes the following contributions:
\begin{itemize}
    \item \rev{We present an exploratory analysis of \cisco vulnerability data and the \bigvul dataset, demonstrating overlapping CVSS distributions and aligned correlations with Kenna‑adjusted scores, which justifies using \bigvul as an open proxy for industrial severity modeling.
    \item We design and evaluate a family of \textit{in‑context configurations} ($C_1$–$C_3$) for CVSS regression, showing that adding a small number of examples and restricting the output format are the key to reducing MSE, non-‐ feasible responses, and scores out ‐ of‑range between models.}
    \item We empirically compare \noncloud and cloud LLMs for CVSS prediction and find that medium‑sized, locally deployable code models—particularly \codellamaS—offer a practical trade‑off between accuracy and privacy, approximating cloud‑model performance while keeping sensitive artifacts on‑premise.
    \item \rev{Finally, we built a \bigvul based testbed and shared our complete empirical evaluation artifacts to facilitate reproductibility\cite{rodriguezcardenas2025artifact}.}

\end{itemize}

\section{Background}\label{sec:background}

\begin{figure*}[ht]
    \centering
    \includegraphics[width=\linewidth]{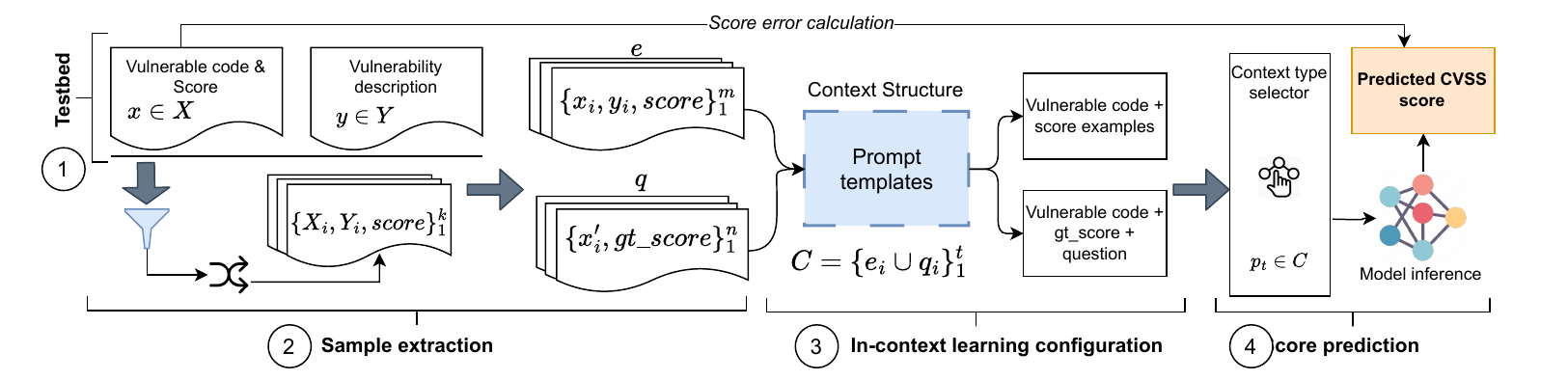} 
    
    \caption{In-Context Learning Solution as a Severity Score Prediction Pipeline}
    \label{fig:overview}
    \vspace{-0.3cm}
\end{figure*}

The Common Vulnerability Scoring System (CVSS) is a standardized scoring method to assess the severity of a computer system security vulnerability. CVSS v3 comprises three base metric groups~\cite{CVSS_UserGuide}. The first group is based on exploitability metrics that include the attack vector, the complexity of the attack, the required privileges, and user interaction. The second group is the impact metric, which includes confidentiality, integrity, and availability. The third group is the scope, designed to capture the impact of the vulnerable component on its linked components. CVSS is the main score industry vendors (\eg Microsoft, Google, Apple, Linux) use to report and prioritize vulnerabilities. We focus on predicting the CVSS score due to its widespread use in industry and association with public datasets. Alternatives such as stakeholder-specific vulnerability categorization (SSVC) and vulnerability priority rating (VPR) incorporate additional factors for severity estimation. SSVC uses a decision tree model for prioritization \cite{spring_2019}, while VPR adds tracking factors and vulnerability history to CVSS V3 \cite{Tenable2023}.


The National Vulnerability Database(NVD) is a comprehensive cybersecurity vulnerability database maintained by the National Institute of Standards and Technology (NIST), the U.S. Department of Commerce agency. NVD serves as the U.S. government repository of standards-based vulnerability management data and acts as the major publisher of the CVSS scores for almost all known vulnerabilities.

The CVSS score is focused on the severity of the vulnerability, new metrics have been proposed to bring more information, for example, the Exploit Prediction Scoring System (EPSS) introduced by Jacobs \etal \cite{jacobs_exploit_2021}, aims to predict the probability that a vulnerability will be exploited in the wild. It provides a probability between 0 and 1, indicating the chance of a vulnerability being exploited within the next 30 days. EPSS, in its first version, uses a logistic regression model trained on 16 features to predict the probability of exploitation within the first year of a vulnerability's disclosure. For the second and third versions of EPSS the logistic regression model switched to a centralized architecture using XGBoost \cite{chen2016xgboost} improving predictive performance.

The Common Vulnerability and Exposures (CVE), found in 1999, is a dictionary of vulnerabilities that have been identified in various code bases \cite{mann1999towards}. Each vulnerability is associated with a unique identifier called CVE ID. The CVE program is maintained by MITRE corporation and sponsored by the U.S. Department of Homeland Security(DHS) and the Cybersecurity and Infrastructure Security Agency(CISA).
Efforts have proposed using severity to calculate a vulnerability's exploitability using heuristics and context aggregation (\eg social media) \cite{bozorgi_beyond_2010, sabottke_vulnerability_nodate,fang_fastembed_2020,xiao_patching_nodate}. Cisco's Kenna risk score evaluates vulnerabilities based on organizational risk, using machine learning for dynamic scoring. The Exploit Prediction Scoring System (EPSS) predicts a vulnerability's exploitation likelihood within 30 days. Both Kenna and EPSS use CVSS scores to predict exploitability. This report evaluates \llms with CVSS, noting their time-series forecasting limits \cite{tan2024languagemodelsactuallyuseful}. Other \llms categorize vulnerabilities and fix code \cite{li_vuldeepecker_2018, russell_automated_2018,hanif_vulberta_2022,hoque_improved_2021}.

\rev{\textbf{Vulnerability Severity Life Cycle.} \textit{Vulnerability management} focuses on detecting, triaging, and enumerating mitigation action. Therefore, security analysts take into account several dates in the vulnerability life cycle. 1) A new software feature is added or modified, 2) the new feature is deployed, 3) a vulnerability is identified, and then, depending on the severity, a fix is required sooner or later. As a consequence, 4.1) a new fix is released and/or 4.2) an attacker exploits the vulnerability. Recent reports indicate a trend in the reduction time from vulnerability detection and attacker exploit, which means that early detection and assisting security analysis is crucial, especially in safety-critical software. }    


\rev{\textbf{In-Context Learning for Vulnerability Severity.} \label{sub:incontext}
Due to the expensive pre-training and fine-tuning process for adapting \llms to a specific task, in-context learning emerges as a strategy for guiding the model to find an accurate answer\cite{geng_large_2024, luo_-context_nodate}. in-context learning consists of designing a set of interactions using input prompt templates\cite{liu_few-shot_2022}. In the realm of vulnerability prediction, in-context Learning emerges as a pivotal strategy to bolster model efficacy and adaptability. By embracing in-context learning, \llms grasp nuanced contextual cues inherent in vulnerability data, thus fortifying their predictive capabilities and adaptability to varying threat landscapes \cite{fu_chatgpt_2023,liu_software_2023}. Our approach aims to apply in-context learning to \noncloud models that can be executed on the customer infrastructure, as previous studies applied to cloud solutions. }
\section{Approach}\label{sec:approach}

\textbf{Problem Delimitation.} Any change or update in software systems produces a new feature that must be analyzed using severity score systems (\ie CVSS) to assess the risks of releasing such a feature and preventing security threats (\eg unauthorized access, data leakage, and system shutdown). However, predicting the severity of the vulnerability is challenging because security analysts must assess \textit{the impact of a vulnerability within a limited time frame}. In addition, analysis becomes crucial and complex as the number of features and vulnerabilities increases. Therefore, a solution is to help the security analyst with tools that help to assess the severity impact and intervene in the vulnerability life cycle in earlier stages (\ie code development when a new feature is added or modified).

This paper proposes the analysis of \rev{\noncloud models to assist analysts in the prioritization task by predicting the severity score.\noncloud models are particularly interesting for industry, as they can run locally, allowing software companies to maintain control over the environment while avoiding sending sensitive data to third-party AI cloud providers.}

\rev{Several studies mainly focused on vulnerability classification (\ie whether the code snippet is vulnerable or not)~\cite{fu_linevul_2022,li_vulnerability_2021, dl_vuln_survey2024}. In our approach,} the prediction of the severity score is considered as a regression problem instead of a classification problem (\eg none [0], low [0.1-3.9], medium [4-6.9], high [7-8.9] and critical [9-10]). Predicting the severity score as a numeric value rather than a categorical label enables the security analyst to quantify severity and \rev{provides better insights into the model's precision and calibration ---the difference between the predicted and ground-truth distribution--- on vulnerability impact\cite{huang2024calibratinglongformgenerationslarge}}. Predicting just the category level conceals the severity of a given vulnerability. For example, consider that an analyst can only predict severity categories. A predicted category severity of \textit{medium} was obtained. The severity of the regression was actually a hidden/unknown value of $~6.9$. In this scenario, the security analyst failed to determine whether the severity value was close to the low or high categories, reducing the precision of the prediction. \rev{This precision is crucial for prompt remediation of the vulnerability and for reasoning about the issue. We hypothesize that this model calibration benefits from our in-context learning solution since severity scores are directly predicted. }





\textbf{In-Context Learning Solution.} We aim to help security analysts triage software features according to the severity of the vulnerability predicted by in-context learning configurations. We require a set of \noncloud models to implement the solution. 

\textit{Step \circled{1}: Testbed.} To evaluate the performance of these models in predicting vulnerability severity, we first collected a set of vulnerable snippets $x \in X $ with their corresponding scores \rev{(\ie CVSS, EPSS, Kenna Risk)} and description $y \in Y$ (see \figref{fig:overview}). These snippets were collected from proprietary  and open source repositories (refer to \bigvul in \secref{sec:setup}). Our experience report only presents statistical analysis with open source snippets due to privacy agreements. 

\textit{Step \circled{2}: Sample extraction} The second step comprises sample extraction by filtering the code with more than 50 \rev{characters and up to 100 characters (\short size sequence), and code with more than 100 characters and up to 300 characters (\medium size sequence)} with a valid description and CVSS score. Including a snippet of more than 50 characters ensures it includes at least the method signature and sufficient context. At the end of this step, we obtain the two subsets of triples $\{x_i, y_i, score\}$. \rev{Each subset is then used to create both a test dataset $M=\{x_i, y_i, score\}$ and a ground-truth dataset $N=\{x_i, y_i, gt\_score\}$. We created the prompt \emph{examples} for the in-context learning setup, using the $M$ dataset, and the \emph{questions} using the $N$ dataset (\figref{fig:prompt-types}). We always hide the ground-truth score ($gt\_score$) from $N$ to the model.} A question will never be used as an example (\secref{sec:setup}). 
\begin{figure*}[h]
    \centering
    \includegraphics[width=0.8\linewidth]{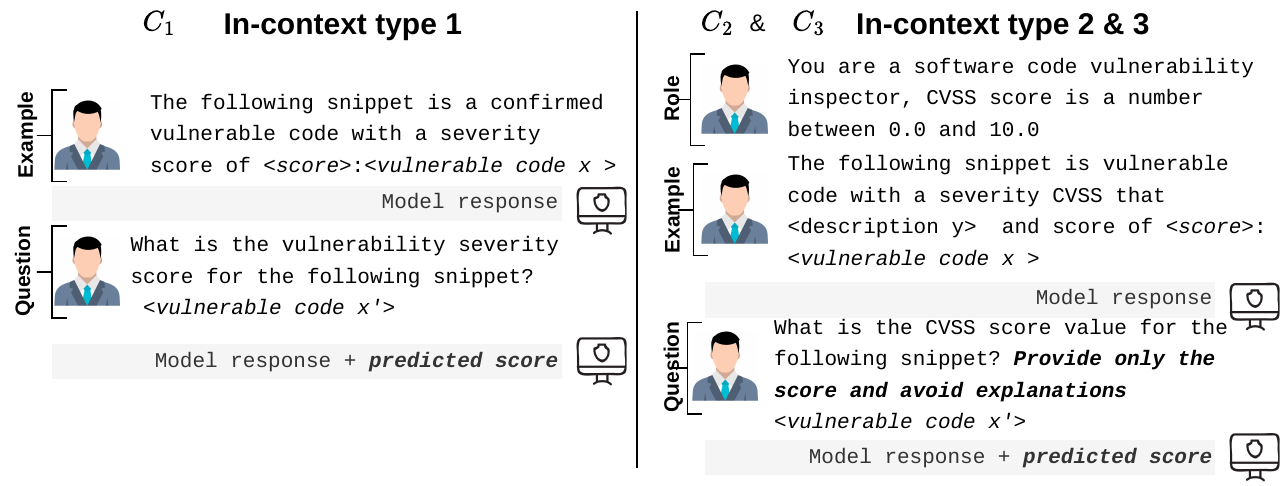}
    \caption{In-Context Learning Configuration. $C_3$ aggregates a specific output format to answer the question compared to $C_2$.}
    \label{fig:prompt-types}
   
\end{figure*}
\textit{Step \circled{3}: In-context Learning Configuration.} The third step presents the testbed generation using a set of prompt templates following a context-type structure (\figref{fig:prompt-types}). A context-type structure concatenates the \emph{example} and the \emph{question}. We propose two context types(\secref{secsec:rq}). These context types require the vulnerable examples $m$ with their current scores and the questions $n$ to complete each prompt, for example $e$ and each prompt for question $q$, respectively. For example, a question prompt template completes the sentence with the vulnerable code $x'_i$ and question $q$.

\textit{Step \circled{4}: Prediction of scores} The fourth step comprises the \textit{context type selector} and the number of examples of shots $l \subset m_p\mid p \in C_t$ to be used. Therefore, the complete context interaction contains the $l$ shot examples of prompt $p\in C_t$ type $t$ with their vulnerable code $x_i$ and the associated score, and the vulnerable code $x'_i$. We conceal the ground truth score \textit{gt\_score} of $x'_i$ from the model. For the context type $C_2$, we aggregate the code description in a context $y_i$. The experiments are executed on each \llm then we capture the predicted score value with a regex expression.
\section{Methodology}\label{sec:methodology}

This section outlines the methodology employed to validate our interpretability technique, we conducted two case studies on xxx popular architectures to explore the following RQs:

\section{Research Questions}\label{secsec:rq}
\rev{In our experience report, we are interested in evaluating the datasets used in the industry and open datasets for vulnerability, the effectiveness of the in-context learning configuration, and the evaluation of \noncloud \llms. 
\begin{enumerate}[label=\textbf{RQ$_{\arabic*}$}, ref=RQ$_{\arabic*}$, wide,labelindent=5pt]\setlength{\itemsep}{0.2em}
    \item \label{rq:data_analysis} \textit{What type of correlations and distributions are there in the industry dataset under study?} We aim to validate the similarity between the industry datasets used and the open-source datasets used to evaluate LLMs for vulnerability detection. The distribution shape allows us to understand the boundaries and limitations for building the in-context approach when using \noncloud models. In addition, because of limitations in reporting \cisco datasets, distribution similarity serves as a proxy assess whether our evaluation generalizes equally to both industry and open-source settings.
    \item\label{rq:best_config}  \textit{Which is the most effective in-context learning configuration to predict vulnerability severity scores?}  We want to detect the most effective \textit{in-context configuration} that enables \llms to improve prediction performance. \figref{fig:prompt-types} illustrates two context templates we fed to our \llms. The first context type $C_1$ is structured using two prompts. The first prompt contains an example of vulnerable code $x$ along with its vulnerability severity \texttt{<score>}, while the second prompt introduces vulnerable code $x'$ asking for severity score. Notice that $C_1$ formulates the examples and questions about the vulnerability severity score without additional information (\eg CVSS concept, ranges of scores, and description of vulnerability).
    The second context type $C_2$ comprises three prompts. The first prompt introduces the models' role, which is a code vulnerable inspector, and the range of the CVSS. This role aggregates more context to the prompt when no examples are available (\ie zero-shot experiments). The second prompt concatenates the vulnerability description \texttt{y}, the vulnerable code example $x$, and the associated ground truth \texttt{<score>}. Finally, the third prompt formulates a question specifically asking for the CVSS score of the corresponding vulnerable code $x'$.
    \item \label{rq:accuracy}  \textit{ How accurate are  \noncloud locally deployed \llms at predicting vulnerability severity scores?} We are interested in evaluating the severity prediction accuracy as a regression problem. Predicting a score value is more informative and precise than classifying the vulnerability by severity ranges, since the score value represents specific submetrics (\secref{sec:background}) that provide complementary information for the security analyst. We aim to support security analysts in selecting the correct \noncloud model by monitoring the \textit{mean square error}.  
\end{enumerate}}

\section{Case Study Design}\label{sec:study_design}

\rev{In this section, we describe the experimental design of our \emph{in-context learning} solution to evaluate \noncloud \llms to predict severity scores.} 

\begin{figure*}[t]
\centering
\begin{subfigure}[t]{0.49\textwidth}
    \centering
    \includegraphics[width=\linewidth]{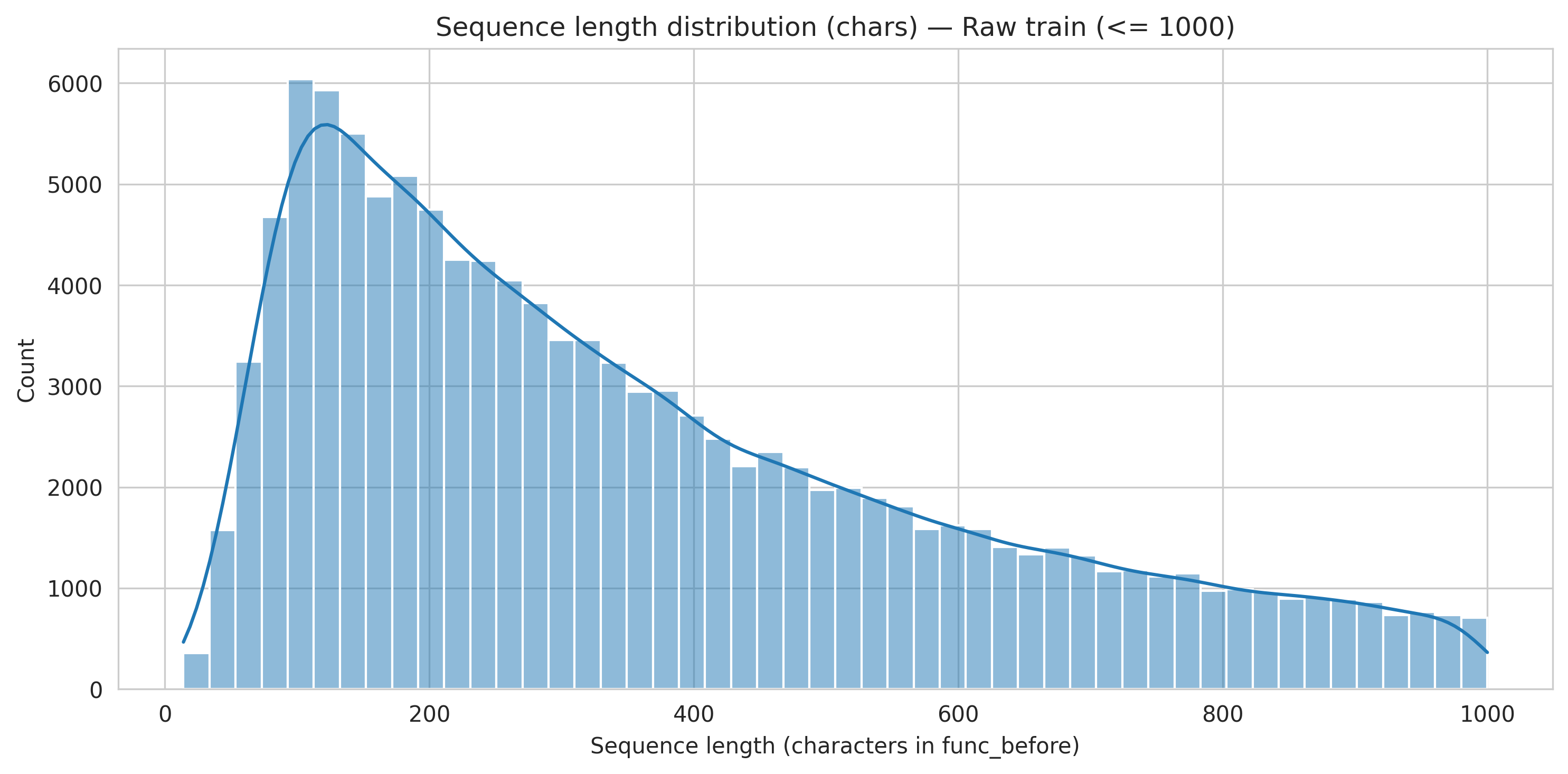}
    \label{fig:bigvul-sub1}
\end{subfigure}
\hfill
\begin{subfigure}[t]{0.49\textwidth}
    \centering
    \includegraphics[width=\linewidth]{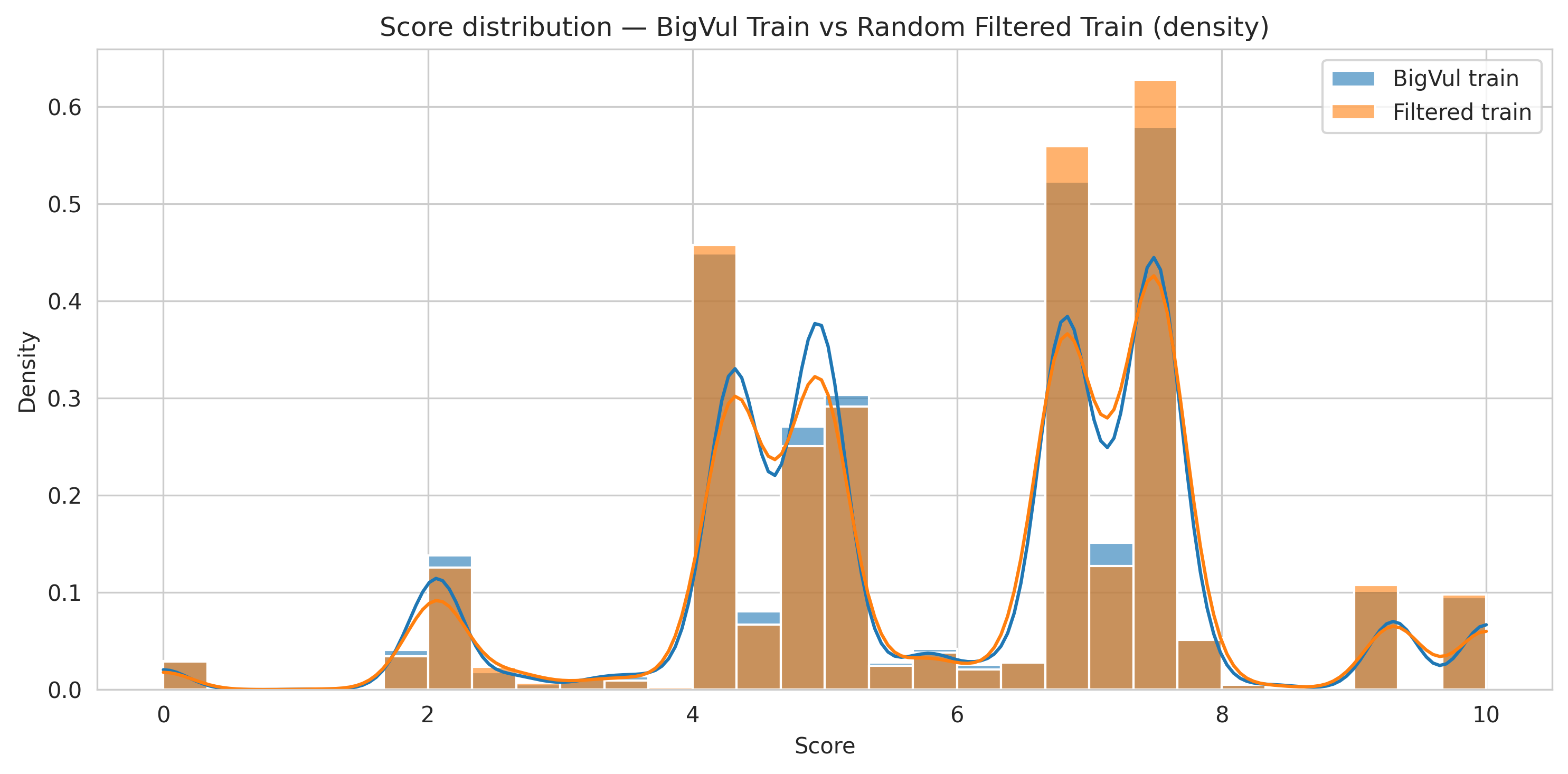}
    \label{fig:bigvul-sub2}
\end{subfigure}
\caption{\rev{Big-Vul dataset: on the left, the sequence size frequency; on the right, the CVSS score distribution for raw data and filtered dataset. }}
\label{fig:bugvul}
\end{figure*}
\subsection{Setup}\label{sec:setup}
\textbf{Testbeds.} {To build our \textit{in-context learning configuration}, we extracted vulnerable code, scores, and description triplets from \bigvul\cite{fan_cc_2020}. \bigvul is a dataset with $\approx 150k$ data points. We filter \bigvul following the second step of our approach (\secref{sec:approach}). \bigvul already includes a curated dataset with the vulnerable code, description, and score. We compare \bigvul to a proprietary dataset which contains many additional variables (\ie EPSS, number of bugs, Kenna risk score). The proprietary dataset includes variables such as whether there are existing malware exploits and how extensively this vulnerability is discussed to help assess impact and risk. The proprietary data provide us with additional information and context for understanding the impact of a bug. 

\rev{After splitting \bigvul into the two datasets, we obtained $\approx 53K$ $m$ triplets for the \textit{example} generation and $\approx 5K$ $n$ triplets for \textit{question} generation (see Table~\ref{tab:big-vul-dataset}). We further filtered the \textit{bigvul} dataset to include only sequences of 50-300 characters. The lower bound of 50 characters ensures that each sequence contains at least the function definition, while the upper bound of 300 characters balances the large number of experimental configurations against the computational cost of inference. Additionally, as shown in Figure~\ref{fig:bugvul}, the majority of the data points fall within this range, and the CVSS score distribution remains the same proportion.}

\rev{From this filtered dataset, we built the example and question prompts by randomly selecting $m$ vulnerable code from the \textit{example} dataset and selecting $n$ vulnerable code from the \textit{question} dataset.   For example, for an experiment of $l=3-shots$, we selected code examples of $m=900$ and $n=300$ questions. We generate the set of interactions for each context type $p\in C_t$ by concatenating each example with its corresponding question. We selected a maximum of $l=3$ due to the time-consuming nature of testing the model with additional examples and the input context size.} 

\rev{Our \ref{rq:best_config} and \ref{rq:accuracy} require two testbeds,  the first testbed uses stratified CVSS classification, and the second uses random CVSS extraction. For the stratified setting, we filtered data points with a sequence length between 50 and 300 characters. Then we classified both trained and test splits according to CVSS severity impact levels: none [0], low [0.1-3.9], medium [4-6.9], high [7-8.9], and critical [9-10](see Tab.~\ref{tab:severity_counts}). The second testbed consists of a randomized sample drawn from the filtered dataset, including both questions and examples. This testbed is further partitioned by sequence length into two groups—50–100 (\short sequence size) and 100–300 characters  (\medium sequence size)—to evaluate the impact of sequence size on performance.}

\begin{table}[h]
\centering
\caption{Open-data \bigvul size after curation} 
\label{tab:big-vul-dataset}

\scalebox{0.9}{%
\setlength{\tabcolsep}{5pt} 
\begin{tabular}{llcc}
 &  & \textbf{Training} & \textbf{Test} \\ \hline
\multicolumn{1}{c}{\multirow{2}{*}{\textit{\textbf{Sequence Size}}}} & \medium & 43856 & 5524 \\
\multicolumn{1}{c}{} & \short & 10425 & 1334 \\
\multicolumn{1}{c}{\textit{\textbf{Documentation}}} & \begin{tabular}[c]{@{}l@{}}Snippet +\\ Summary\end{tabular} & 139607 & 17568 \\ \hline
 & \multicolumn{1}{c}{\textbf{Total}} & 150908 & 18864 \\ \bottomrule
\end{tabular}
}
\end{table}
\begin{table}[ht]
\centering
\caption{Counts per severity impact for train and test datasets with combined sequence size}
\begin{tabular}{lrr}
\textbf{Severity} & \textbf{Train Count} & \textbf{Test Count} \\
\hline
None (0)        & 530   & 63  \\
Low (0.1--3.9)  & 3853  & 472 \\
Medium (4--6.9) & 31386 & 3926 \\
High (7--8.9)   & 14749 & 1889 \\
Critical (9--10)& 3763  & 508 \\
\hline
\end{tabular}

\label{tab:severity_counts}
\end{table}
\noindent \textbf{Machine Configuration.} We performed the experiments using 20.04 Ubuntu with an AMD EPYC 7532 32-core CPU, an A100 NVIDIA GPU with 40GB VRAM, and 1TB RAM. For the model inference process, we used HuggingFace and Pytorch \cite{wolf2020transformers, pytorch}. All models were loaded into the GPU to increase the inference time.

\noindent \textbf{Open-source \llms for Code.} Our experiments utilized three types of \llms. Firstly, \textit{CodeLlama2}, is a specialized language model for programming tasks developed by Meta. The model can effectively identify potential security flaws and vulnerabilities in code by recognizing patterns indicative of weaknesses. The model also demonstrates proficiency in code interpretation, debugging, and documentation generation~\cite{touvron2023llama}. We used the Huggingface 7B and 13B model sizes for our experiments. Secondly, \textit{Mistral}, is an open model that includes code generation capabilities reporting even better performance than Llama2\cite{mistral2023}. \rev{We used the 7B-size versions of these models via the HuggingFace \texttt{transformers} library and ran Mixtral through the Ollama runtime for our experiments~\cite{wolf2020transformers,hf_mixtral_docs,ollama_docs,mixtral_ollama}. Finally, we evaluated OpenAI GPT4o-mini and GPT-OSS models~\cite{openai_gpt4omini,ollama_gptoss}. GPT4o-mini is our reference cloud model; GPT-OSS is an open-weight model that works with Ollama in a local environment.  For all experiments, we set the following parameters: temperature of 0.001, 0.3, and 0.9, top-k of 50, and top-p of 0.92 for CodeLlama and Mistral models and default parameters for OpenAI models.} 

\rev{\noindent \textbf{In-context configurations.} Figure~\ref{fig:prompt-types} illustrates the in-context setups used in our experiments. Following \textit{Step \circled{4}} in Section~\ref{sec:approach}, we define three types of context. The first configuration, $C_1$, consists of a set of examples determined by the number of shots, along with a direct query regarding a vulnerable code instance $x'$.
The second configuration, $C_2$, extends this setup by incorporating a system prompt that assigns the model the role of a vulnerability inspector, followed by the examples and the query.
The third configuration, $C_3$, further enhances the prompt by specifying the required output format in addition to the system prompt and examples. This design restricts the response to only the predicted CVSS score, preventing the inclusion of nonessential information.}

\subsection{Statistical Evaluation}
{We configured two experiments to predict the vulnerability severity score. \rev{The first is the stratified experiment, which consists of $600$ context samples, each context sample with 3 shots. The second experiment is a random sample experiment that aims to evaluate the CVSS score prediction as a regular vulnerability inspector with a spontaneous snippet of code. Each experiment consists of $300$ context samples. Each context sample picks a random question to ask the severity score; the number of examples that relate to the question depends on the number of shots (\secref{secsec:rq}). For experiments using CodeLlama or Mistral with temperature 0.3 or 0.9, we collect the model response range by running each experiment $30$ times. We aim to capture the model diversity outcome and also validate the default model's parameters. Once the experiment is completed, we apply the following statistical methodology to evaluate our AI solution in terms of \textit{in-context configuration effectiveness} and model accuracy. \textit{Step 1:} Calculate the square difference error between predicted and expected score (\textit{gt\_score}) for each trial \textit{Step 2:} Replace \textit{unfeasible} values with the median of feasible predicted severity scores. An unfeasible value indicates that the generated text does not resemble an accepted severity value. Instead of predicting a concrete value, the model only generates a description of the code from the question prompt (\ie \texttt{'Based on the provided code snippet alone, it is not possible or accurate to determine a CVSS score...'}). \textit{Step 3:} Compute the mean square error (\metricMSE) for the 30 trials for each sample. To evaluate accuracy and in-context configuration effects, we compute the \textit{ mean MSE} among the 300 samples.}
In addition, we computed the following metrics. \metricNonFeasible constitutes the proportion of non-feasible or unanswered samples {from the total of samples (\ie $\sum{unfeasible/ n}$)}.  \metricOutofRange embodies the number of trials when the model predicted a severity score outside the CVSS range (\ie predicted score $> 10.0$ or $<0.0$). \metricExMean and \metricExStd are the mean and standard deviation of the score values in the input prompt examples. Finally, the \metricGTMean and \metricGTStd are the mean and standard deviation of the expected severity score (\ie ground truth scores).

{ \ref{rq:best_config} and \ref{rq:accuracy} employed all the previous metrics. However, to answer \ref{rq:best_config} we variated the context $C$ to compare the performance, meanwhile, we focused on sequence length, and model variation to answer \ref{rq:accuracy}.} \rev{The sequence lenght is calculated by the number of tokens used at the input prompt.} 


\section{RQ$_1$ Exploratory Data Analysis}\label{sec:eda}

This section outlines the two datasets used in our case study, which serve as the basis for building prompts to evaluate the \noncloud models and configure test scenarios addressing our research questions. Since the proprietary dataset is confidential and has a limited number of data points, we supplemented it with an open-source dataset of vulnerabilities called \bigvul.

We conducted two types of analysis on the \cisco dataset and \bigvul: a distribution comparison and an evaluation of the correlations between the datasets. These analyses aim to assess the naturalness of both datasets in generating our prompt testbed. By examining naturalness, we can determine whether the open-source dataset contains information comparable to the proprietary dataset, allowing us to use it interchangeably if needed.

\subsection{Distribution Comparison}
\begin{figure*}[t]
\centering
\begin{subfigure}[t]{0.49\textwidth}
    \centering
    \includegraphics[width=\linewidth]{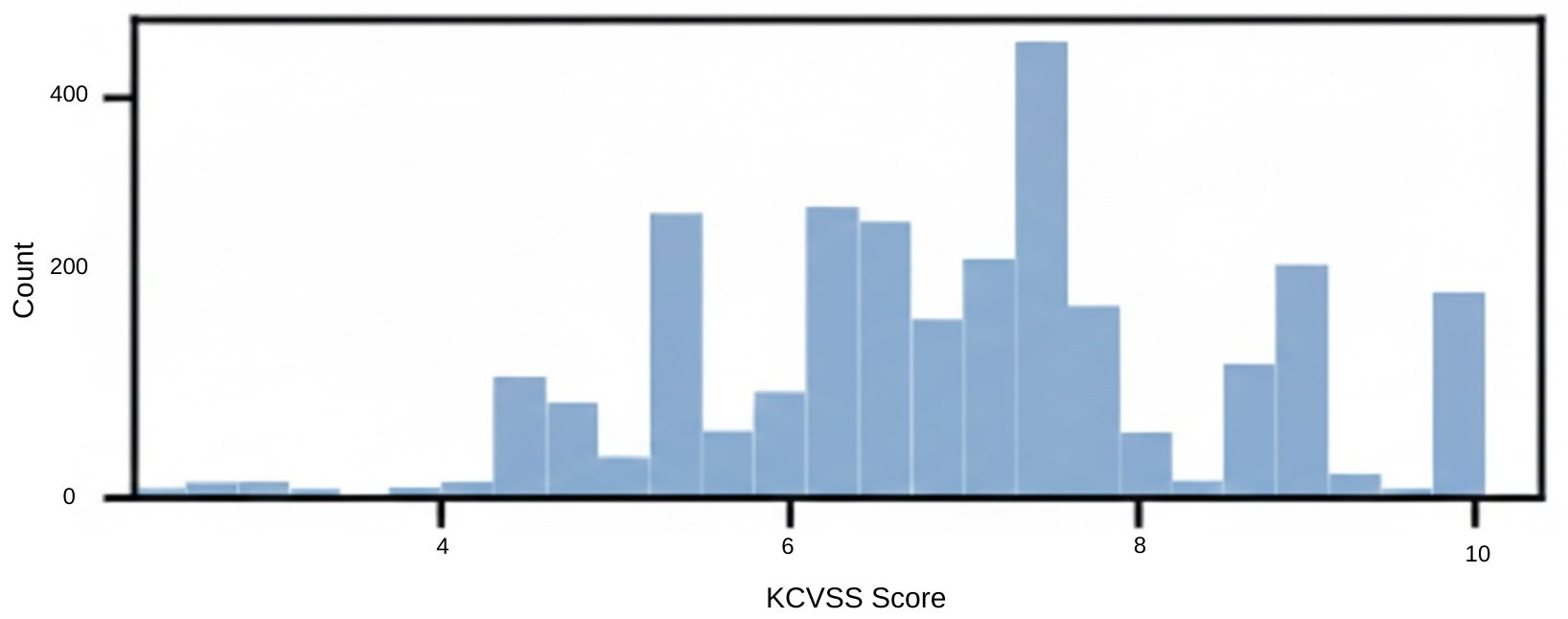}
    \label{fig:kcvss-sub1}
\end{subfigure}
\hfill
\begin{subfigure}[t]{0.49\textwidth}
    \centering
    \includegraphics[width=\linewidth]{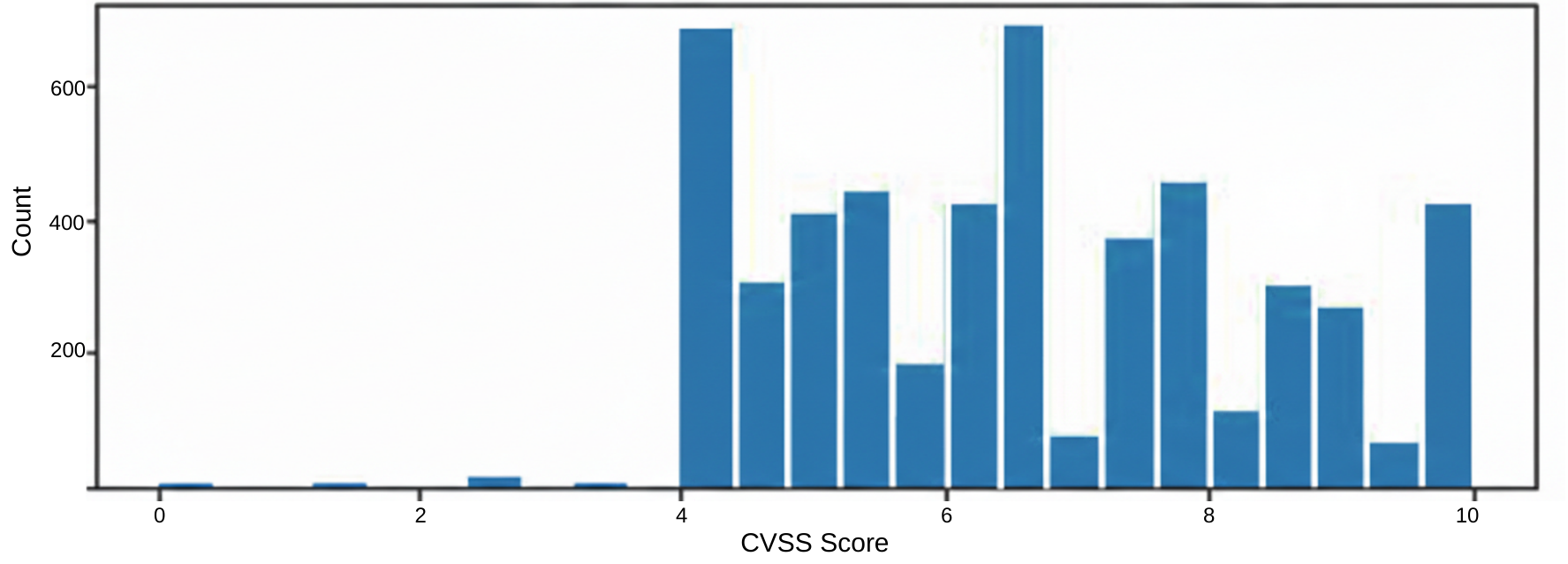}
    \label{fig:kcvss-sub2}
\end{subfigure}

\caption{\rev{Proprietary dataset frequencies: on the left, the KCVSS score, on the right, the assigned CVSS to the proprietary snippets.}}
\label{fig:cisco_kcvss_freq}
\end{figure*}

\textbf{Industry Dataset Shape: }
The \cisco data comprises three distinct datasets, each contributing valuable insights into vulnerabilities and risk assessment. The first dataset is the PSIRT CVE Vulnerability Dataset, which contains 5,520 data points. This data set includes detailed information such as an identifier (ID), associated CVE (Common Vulnerabilities and Exposures) and associated BugIDs. It also provides a title and description of each vulnerability, along with its severity rating, CVSS (Common Vulnerability Scoring System) score, and KCVSS (Kenna-adjusted CVSS) score. \rev{\cisco dateset includes additional fields associated with the CWEs (Common Weakness Enumerations) and products affected by each vulnerability.}

The second dataset is the Kenna Risk Dataset, comprising 5,372 data points. This dataset emphasizes risk factors and associated metadata for vulnerabilities. Key attributes include the CVE identifier and the Kenna Risk Score, which quantify risk. It also captures chatter indicators, reflecting the frequency of online discussions, and tags vulnerabilities for characteristics such as malware, remote code execution, and ease of exploitation. Additional fields flag vulnerabilities with active malware exploitation, breaches, and discussions that occur before the publication of the National Vulnerability Database (NVD).

The third dataset, the EPSS Dataset, is the largest, containing 221,346 data points. It focuses on predictive metrics and ranking of vulnerabilities. Key attributes include the CVE identifier, the EPSS (Exploit Prediction Scoring System) score—which estimates the likelihood of a vulnerability being exploited—and percentile rankings that position vulnerabilities within a broader exploitability landscape.

These datasets span 13 years, from 2010 to 2023, allowing a comprehensive longitudinal analysis of trends in vulnerability and risk. To provide additional context and validation, we compare the \rev{\cisco data against the publicly available BigVul dataset, a well-known resource in vulnerability research\cite{fan_cc_2020}.}

Through this comparison, our objective is to uncover patterns, correlations, and insights that deepen our understanding of vulnerability, naturalness, risk prioritization, and precision. Together, these datasets form a solid foundation for our analysis, offering unique information dimensions critical to addressing our research objectives.

Fig.~\ref{fig:cisco_kcvss_freq} shows the number of records that have each CVSS score for the \cisco and \bigvul datasets. Very few records have a CVSS score below 4. Compared to the \cisco dataset, the \bigvul dataset contains fewer records with high CVSS scores. However, the majority of CVSS scores in both datasets fall in the range of 4 to 8. The \cisco dataset contains decimals, as shown in \figref{fig:cisco_kcvss_freq}. Since \bigvul has a larger number of data points than \cisco, we group them by integer values. We notice that \bigvul has a higher frequency on the CVSS score of 7 than \cisco, however, the distribution overlaps and indicates a similar distribution.

\subsection{Distribution Correlation}

\figref{fig:cisco_cvss_kenna_heatmap} shows a heatmap of the relationship between the CVSS and \textit{KCVSS} scores. The color of each box indicates the number of records with the respective CVSS and KCVSS scores. The heatmap on the right at \figref{fig:cisco_cvss_kenna_heatmap} indicates a correlation between CVSS and KCVSS. KCVSS and CVSS are scores between 0 and 10.0, with the highest frequency at score 6. The heatmap on the right indicates a different correlation between \textit{Kenna Risk} and CVSS. 

    \begin{figure}[htbp]
    \begin{subfigure}[b]{0.235\textwidth}
        \centering
    \includegraphics[width=\linewidth]{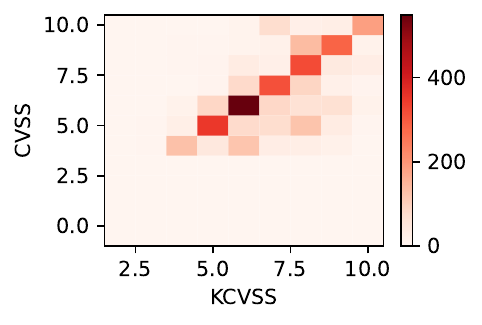}
   
    \end{subfigure}
    \begin{subfigure}[b]{0.235\textwidth}
        \centering
      \includegraphics[width=\linewidth]{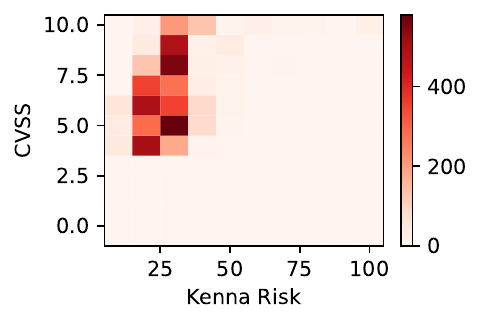}
   
    \end{subfigure}
    \caption{\cisco CVSS correlation with KCVSS and kenna risk, color represents the frequency.}
      \label{fig:cisco_cvss_kenna_heatmap}
      \vspace{-0.8em}
\end{figure}

\rev{\textbf{Discussion.} Comparing this \cisco data to the publicly available \bigvul dataset reveals overlapping CVSS score distributions, with notable alignment in the 4–8 range but variations such as \bigvul's higher frequency of CVSS 7 scores.} This robust dataset suite provides nuanced insights into vulnerability patterns, risk prioritization, and exploitability. Although the CVSS-KCVSS relationship shows strong alignment, we observe a weaker correlation between  CVSS and Kenna Risk scores.



\begin{table}[]
\centering
\caption{Mean Squared Error calculated for each model with sequences \short and \medium size combined with 3-shots, varying the temperature $T$} 
\label{tab:temp_test}

\scalebox{0.8}{%
\begin{tabular}{cccccccc}
\toprule
\textbf{Model} &
  \textbf{$C$} &
  \textbf{$T$} &
  \textbf{\begin{tabular}[c]{@{}c@{}}MSE\\ Mean\end{tabular}} &
  \textbf{\begin{tabular}[c]{@{}c@{}}Std.\\ MSE\end{tabular}} &
  \textbf{\begin{tabular}[c]{@{}c@{}}Gt.\\ Mean\end{tabular}} &
  \textbf{\begin{tabular}[c]{@{}c@{}}Gt.\\ Std.\end{tabular}} &
  \textbf{\begin{tabular}[c]{@{}c@{}}\% Non\\ Feasible\end{tabular}} \\ \hline
\multirow{6}{*}{\begin{tabular}[c]{@{}c@{}}CodeLlama2\\ 7B\end{tabular}} &
  \multirow{2}{*}{$C_1$} &
  0.001 &
  7.91 &
  13.01 &
  6.14 &
  1.82 &
  0.00 \\
 &
   &
  0.300 &
  \cellcolor[HTML]{DAE8FC}6.69 &
  \cellcolor[HTML]{DAE8FC}9.63 &
  6.44 &
  1.33 &
  0.00 \\
 &
  \multirow{2}{*}{$C_2$} &
  0.001 &
  8.36 &
  12.74 &
  6.13 &
  1.83 &
  \cellcolor[HTML]{FFCCC9}0.50 \\
 &
   &
  0.300 &
  \cellcolor[HTML]{DAE8FC}6.98 &
  10.35 &
  6.10 &
  1.89 &
  0.00 \\
 &
  \multicolumn{1}{l}{\multirow{2}{*}{$C_3$}} &
  0.001 &
  10.00 &
  16.21 &
  6.06 &
  1.88 &
  4.00 \\
 &
  \multicolumn{1}{l}{} &
  0.300 &
  10.87 &
  15.76 &
  6.21 &
  1.73 &
  4.50 \\ \hline
\multirow{6}{*}{\textit{\begin{tabular}[c]{@{}c@{}}CodeLlama\\ 13B\end{tabular}}} &
  \multirow{2}{*}{$C_1$} &
  0.001 &
  10.25 &
  13.49 &
  6.19 &
  1.89 &
  0.60 \\
 &
   &
  0.300 &
  12.31 &
  16.49 &
  6.11 &
  1.99 &
  \cellcolor[HTML]{FFCCC9}5.00 \\
 &
  \multirow{2}{*}{$C_2$} &
  0.001 &
  27.58 &
  23.87 &
  6.15 &
  1.92 &
  0.23 \\
 &
   &
  0.300 &
  26.42 &
  24.60 &
  6.08 &
  1.94 &
  \cellcolor[HTML]{FFCCC9}0.68 \\
 &
  \multirow{2}{*}{$C_3$} &
  0.001 &
  \cellcolor[HTML]{DAE8FC}9.13 &
  16.67 &
  6.06 &
  1.80 &
  0.00 \\
 &
   &
  0.300 &
  10.81 &
  16.47 &
  6.16 &
  1.98 &
  0.50 \\ \hline
\multirow{6}{*}{\gptoss} &
  \multirow{2}{*}{$C_1$} &
  0.001 &
  32.43 &
  23.77 &
  5.95 &
  1.95 &
  0.85 \\
 &
   &
  0.300 &
  34.65 &
  23.62 &
  6.14 &
  1.89 &
  \cellcolor[HTML]{FFCCC9}2.80 \\
 &
  \multirow{2}{*}{$C_2$} &
  0.001 &
  37.89 &
  23.98 &
  6.06 &
  1.93 &
  0.10 \\
 &
   &
  0.300 &
  35.39 &
  22.82 &
  5.88 &
  1.88 &
  0.90 \\
 &
  \multirow{2}{*}{$C_3$} &
  0.001 &
  34.91 &
  23.84 &
  6.09 &
  1.90 &
  2.10 \\
 &
   &
  0.300 &
  37.28 &
  24.86 &
  6.06 &
  2.07 &
  2.00 \\ \bottomrule
\end{tabular}%
}
{\\ \footnotesize *Best \metricMSE in blue. Worst non-feasible proportions in red.}
\label{tab:my-table}
\end{table}
\section{RQ$_2$ In-context Configuration Effectiveness}\label{sec:effectiveness}
The purpose of evaluating the \textit{in-context configuration} for predicting vulnerability severity is to observe how the model performs when we provide additional context explaining its role as a vulnerability evaluator using \rev{the context samples $C_1$, $C_2$, and $C_3$. We started by testing the \noncloud model's temperature impact, then varied the context and the number of shots.}

\tabref{tab:temp_test} depicts the \textit{Avg.} \metricMSE of the severity score prediction and the standard deviation for each type of context $C$ \rev{for \codellama, \mistral, and \gptoss with sequences from 50 to 300 characters. \codellamaS outperforms other models, reaching the lowest \metricMSE (6.69) and a 0.0\% non-feasible rate. While \codellamaL performs similarly, its non-feasible rate degrades significantly at a higher temperature of 0.3 compared to its deterministic 0.001 setting. \gptoss consistently underperforms across all conditions, with error rates roughly four to five times higher than the top-performing \codellama configuration. We observed similar values for the Mixtral model with \gptoss with the largest \metricMSE values (see Appendix~\cite{rodriguezcardenas2025artifact}).}

\begin{table}[]
\centering
\caption{Mean Squared Error calculated for $C_1$,$C_2$, and $C_3$ among shots $i=0,1,3$ and temperature $T=0.3$} 
\label{tab:mse-03}

\scalebox{0.70}{%
\begin{tabular}{ccccccccccc}
\toprule
\textbf{Model} &
  \textbf{$i$} &
  \textbf{$C$} &
  \textbf{\begin{tabular}[c]{@{}c@{}}Seq. \\ Size\end{tabular}} &
  \textbf{\begin{tabular}[c]{@{}c@{}}MSE \\ Mean\end{tabular}} &
  \textbf{\begin{tabular}[c]{@{}c@{}}Std. \\ MSE\end{tabular}} &
  \textbf{\begin{tabular}[c]{@{}c@{}}\% \textit{Non} \\ \textbf{Feasible}\end{tabular}} &
  \textbf{\begin{tabular}[c]{@{}c@{}}Gt. \\ Mean\end{tabular}} &
  \textbf{\begin{tabular}[c]{@{}c@{}}Gt. \\ Std.\end{tabular}} &
  \textbf{\begin{tabular}[c]{@{}c@{}}Ex. \\ Mean\end{tabular}} &
  \textbf{\begin{tabular}[c]{@{}c@{}}Ex. \\ Std.\end{tabular}} \\ \hline
 &
   &
   &
  \short &
  18.46 &
  20.48 &
  0.34 &
  6.36 &
  1.79 &
  - &
  0.00 \\
 &
   &
  \multirow{-2}{*}{$C_1$} &
  \medium &
  11.38 &
  14.93 &
  \cellcolor[HTML]{FFCCC9}\textbf{2.34} &
  5.96 &
  2.00 &
  - &
  0.00 \\
 &
   &
   &
  \short &
  30.16 &
  24.20 &
  0.00 &
  6.36 &
  1.71 &
  - &
  0.00 \\
 &
   &
  \multirow{-2}{*}{$C_2$} &
  \medium &
  14.22 &
  18.17 &
  0.00 &
  6.16 &
  1.71 &
  - &
  0.00 \\
 &
   &
   &
  \short &
  10.18 &
  12.80 &
  0.00 &
  6.47 &
  1.92 &
  - &
  0.00 \\
 &
  \multirow{-6}{*}{0} &
  \multirow{-2}{*}{$C_3$} &
  \medium &
  11.55 &
  12.50 &
  0.00 &
  5.93 &
  1.82 &
  - &
  0.00 \\
 &
   &
   &
  \short &
  8.24 &
  11.68 &
  0.00 &
  6.42 &
  1.72 &
  6.16 &
  0.00 \\
 &
   &
  \multirow{-2}{*}{$C_1$} &
  \medium &
  7.00 &
  10.05 &
  0.34 &
  6.07 &
  1.87 &
  6.00 &
  0.00 \\
 &
   &
   &
  \short &
  8.45 &
  11.97 &
  0.00 &
  6.43 &
  1.90 &
  6.45 &
  0.00 \\
 &
   &
  \multirow{-2}{*}{$C_2$} &
  \medium &
  7.39 &
  9.45 &
  0.00 &
  6.17 &
  1.92 &
  5.91 &
  0.00 \\
 &
   &
   &
  \short &
  6.99 &
  9.36 &
  0.00 &
  6.32 &
  1.75 &
  6.11 &
  0.00 \\
 &
  \multirow{-6}{*}{1} &
  \multirow{-2}{*}{$C_3$} &
  \medium &
  6.96 &
  9.72 &
  0.00 &
  6.28 &
  1.77 &
  6.30 &
  0.00 \\
 &
   &
   &
  \short &
  8.76 &
  12.92 &
  0.67 &
  6.47 &
  1.80 &
  6.16 &
  1.32 \\
 &
   &
  \multirow{-2}{*}{$C_1$} &
  \medium &
  8.94 &
  11.56 &
  1.36 &
  6.09 &
  1.69 &
  6.18 &
  0.00 \\
 &
   &
   &
  \short &
  9.85 &
  13.96 &
  0.00 &
  6.31 &
  1.78 &
  6.16 &
  1.28 \\
 &
   &
  \multirow{-2}{*}{$C_2$} &
  \medium &
  15.49 &
  21.26 &
  0.00 &
  6.01 &
  1.87 &
  6.14 &
  0.00 \\
 &
   &
   &
  \short &
  \cellcolor[HTML]{DAE8FC}\textbf{5.86} &
  7.64 &
  0.00 &
  6.25 &
  1.70 &
  6.19 &
  1.42 \\
\multirow{-18}{*}{\textit{\begin{tabular}[c]{@{}c@{}}CodeLlama2\\ 7B\end{tabular}}} &
  \multirow{-6}{*}{3} &
  \multirow{-2}{*}{$C_3$} &
  \medium &
  12.30 &
  12.83 &
  0.00 &
  5.90 &
  1.76 &
   &
  0.00 \\ \hline
 &
   &
   &
  \short &
  31.63 &
  22.70 &
  1.00 &
  6.26 &
  1.81 &
  - &
  0.00 \\
 &
   &
  \multirow{-2}{*}{$C_1$} &
  \medium &
  21.39 &
  20.00 &
  \cellcolor[HTML]{FFCCC9}\textbf{3.36} &
  5.97 &
  1.77 &
  - &
  0.00 \\
 &
   &
   &
  \short &
  26.90 &
  24.02 &
  0.00 &
  6.25 &
  1.84 &
  - &
  0.00 \\
 &
   &
  \multirow{-2}{*}{$C_2$} &
  \medium &
  25.25 &
  23.67 &
  0.00 &
  6.10 &
  1.76 &
  - &
  0.00 \\
 &
   &
   &
  \short &
  9.91 &
  12.09 &
  0.00 &
  6.05 &
  1.79 &
  - &
  0.00 \\
 &
  \multirow{-6}{*}{0} &
  \multirow{-2}{*}{$C_3$} &
  \medium &
  10.57 &
  12.75 &
  0.00 &
  5.92 &
  1.86 &
  - &
  0.00 \\
 &
   &
   &
  \short &
  13.74 &
  18.33 &
  0.00 &
  6.24 &
  1.87 &
  6.22 &
  0.00 \\
 &
   &
  \multirow{-2}{*}{$C_1$} &
  \medium &
  9.04 &
  12.82 &
  0.00 &
  6.07 &
  1.80 &
  5.80 &
  0.00 \\
 &
   &
   &
  \short &
  19.95 &
  21.38 &
  0.00 &
  6.26 &
  1.89 &
  6.16 &
  0.00 \\
 &
   &
  \multirow{-2}{*}{$C_2$} &
  \medium &
  13.29 &
  18.85 &
  0.00 &
  5.98 &
  1.91 &
  5.93 &
  0.00 \\
 &
   &
   &
  \short &
  5.80 &
  8.04 &
  0.00 &
  6.41 &
  1.73 &
  6.41 &
  0.00 \\
 &
  \multirow{-6}{*}{1} &
  \multirow{-2}{*}{$C_3$} &
  \medium &
  6.07 &
  9.45 &
  0.00 &
  6.20 &
  1.80 &
  5.95 &
  0.00 \\
 &
   &
   &
  \short &
  13.89 &
  18.37 &
  0.00 &
  6.26 &
  1.74 &
  6.14 &
  1.36 \\
 &
   &
  \multirow{-2}{*}{$C_1$} &
  \medium &
  11.04 &
  15.71 &
  0.00 &
  5.84 &
  1.86 &
  5.88 &
  1.38 \\
 &
   &
   &
  \short &
  17.54 &
  21.54 &
  0.00 &
  6.26 &
  1.75 &
  6.21 &
  1.37 \\
 &
   &
  \multirow{-2}{*}{$C_2$} &
  \medium &
  11.31 &
  15.76 &
  0.00 &
  6.06 &
  1.79 &
  5.89 &
  1.41 \\
 &
   &
   &
  \short &
  \cellcolor[HTML]{DAE8FC}\textbf{4.90} &
  6.58 &
  0.00 &
  6.09 &
  1.81 &
  6.16 &
  1.36 \\
\multirow{-18}{*}{\textit{\begin{tabular}[c]{@{}c@{}}CodeLlama2\\ 13B\end{tabular}}} &
  \multirow{-6}{*}{3} &
  \multirow{-2}{*}{$C_3$} &
  \medium &
  6.49 &
  8.28 &
  0.00 &
  5.99 &
  1.97 &
  5.98 &
  1.37 \\ \hline
 &
   &
   &
  \short &
  18.65 &
  22.82 &
  0.00 &
  6.21 &
  1.93 &
  - &
  0.00 \\
 &
   &
  \multirow{-2}{*}{$C_1$} &
  \medium &
  8.62 &
  11.26 &
  0.00 &
  6.02 &
  1.92 &
  - &
  0.00 \\
 &
   &
   &
  \short &
  18.22 &
  20.14 &
  0.00 &
  5.85 &
  1.86 &
  - &
  0.00 \\
 &
   &
  \multirow{-2}{*}{$C_2$} &
  \medium &
  8.95 &
  10.40 &
  0.00 &
  6.11 &
  1.70 &
  - &
  0.00 \\
 &
   &
   &
  \short &
  3.44 &
  4.77 &
  0.00 &
  6.26 &
  1.87 &
  - &
  0.00 \\
 &
  \multirow{-6}{*}{0} &
  \multirow{-2}{*}{$C_3$} &
  \medium &
  3.79 &
  5.67 &
  0.00 &
  6.12 &
  1.85 &
  - &
  0.00 \\
 &
   &
   &
  \short &
  13.79 &
  19.94 &
  0.00 &
  6.29 &
  1.83 &
   &
  0.00 \\
 &
   &
  \multirow{-2}{*}{$C_1$} &
  \medium &
  9.82 &
  14.10 &
  0.00 &
  6.04 &
  1.93 &
  5.93 &
  0.00 \\
 &
   &
   &
  \short &
  11.10 &
  17.92 &
  0.00 &
  6.15 &
  1.86 &
  6.12 &
  0.00 \\
 &
   &
  \multirow{-2}{*}{$C_2$} &
  \medium &
  7.85 &
  10.99 &
  0.00 &
  5.98 &
  1.93 &
  5.92 &
  0.00 \\
 &
   &
   &
  \short &
  3.46 &
  5.38 &
  0.00 &
  6.34 &
  1.77 &
  6.15 &
  0.00 \\
 &
  \multirow{-6}{*}{1} &
  \multirow{-2}{*}{$C_3$} &
  \medium &
  3.67 &
  5.38 &
  0.00 &
  6.01 &
  1.78 &
  5.77 &
  0.00 \\
 &
   &
   &
  \short &
  7.19 &
  9.72 &
  0.00 &
  6.42 &
  1.72 &
  5.92 &
  0.00 \\
 &
   &
  \multirow{-2}{*}{$C_1$} &
  \medium &
  5.85 &
  8.77 &
  0.00 &
  6.02 &
  1.76 &
  6.09 &
  1.34 \\
 &
   &
   &
  \short &
  7.14 &
  10.71 &
  0.00 &
  6.15 &
  1.86 &
  6.11 &
  0.00 \\
 &
   &
  \multirow{-2}{*}{$C_2$} &
  \medium &
  5.41 &
  7.06 &
  0.00 &
  6.09 &
  1.67 &
  5.91 &
  1.35 \\
 &
   &
   &
  \short &
  \cellcolor[HTML]{DAE8FC}\textbf{3.08} &
  4.58 &
  0.00 &
  6.25 &
  1.79 &
  5.89 &
  0.00 \\
\multirow{-18}{*}{\textit{\begin{tabular}[c]{@{}c@{}}GPT-4o\\ mini\end{tabular}}} &
  \multirow{-6}{*}{3} &
  \multirow{-2}{*}{$C_3$} &
  \medium &
  3.17 &
  4.78 &
  0.00 &
  6.07 &
  1.82 &
  5.98 &
  1.40 \\ \hline
\end{tabular}%
}
{\\ \footnotesize *Best \metricMSE in blue-bold. Worst non-feasible proportions in red.}
\end{table}

\rev{The temperature calibration results indicate that \codellamaS achieves better predictive performance at $T=0.3$ while maintaining a low non-feasible rate. Although \codellamaL performs slightly better with $C_3$ at $T=0.001$, its performance at $T=0.3$ remains comparable. Given that default temperature settings for \noncloud models are typically higher than $T=0.001$, we selected a temperature of $0.3$ for our empirical in-context effectiveness evaluation.}

\rev{Table~\ref{tab:mse-03} show a empirical evaluation for our \textit{in-context} evaluation. For all three models, increasing the sequence size from \short to \medium and the number of iterations from $i=0$ to $i=3$ tends to reduce \metricMSE and stabilize performance, with the best trade‑offs between accuracy and feasibility appearing at larger sequences and higher iteration counts. Compared to \codellamaS, which shows the highest \metricMSE and occasional \metricNonFeasible runs, \codellamaL narrows the gap to \gptf but still trails it, confirming \gptf as the strongest model overall in this setting. Table~\ref{tab:mse-03} confirms the observed results of our temperature calibration experiment, and $C_3$ constitutes the best prompt configuration to predict the CVSS score among the \codellama model and our reference model \gptf. Notably, we observe the highest \metricNonFeasible with the \codellamaL model for sequences with \medium size and context $ C_1$. We also observe similar behavior with \codellamaS, with a slightly lower value of $2.34$ for \medium size sequences and $C_1$. We avoided the \mistral model as we observed \metricMSE of $\approx 33$ for all context configurations (see appendix~\cite{rodriguezcardenas2025artifact}).}

\begin{table}[]
\centering
\caption{Mean Squared Error calculated for each model and sequence sizes with temperature of $T=0.9$} 
\label{tab:socre_evaluation}

\scalebox{0.62}{%
\begin{tabular}{cccccccccccc}

\hline
\textbf{Model} & \multicolumn{1}{l}{$C$} & \textbf{\begin{tabular}[c]{@{}c@{}}Seq.\\ Size\end{tabular}} & \textbf{$l$} & \textbf{\begin{tabular}[c]{@{}c@{}} \textit{Avg.} \\ \metricMSE \end{tabular}} & \textbf{\begin{tabular}[c]{@{}c@{}}\textit{Std.}\\ \metricMSE \end{tabular}} & \textbf{\begin{tabular}[c]{@{}c@{}} \textit{Non} \\ \textbf{Feasible}\end{tabular}} & \textbf{\begin{tabular}[c]{@{}c@{}} \textit{Out-Of-}\\ \textit{Range}\end{tabular}} & \textbf{\begin{tabular}[c]{@{}c@{}} \textit{Ex.}\\ \textit{Mean}\end{tabular}} & \textbf{\begin{tabular}[c]{@{}c@{}} \textit{Ex.} \\ \textit{Std.}\end{tabular}} & \textbf{\begin{tabular}[c]{@{}c@{}} \textit{Gt.}\\ \textit{Mean}\end{tabular}} & \textbf{\begin{tabular}[c]{@{}c@{}} \textit{Gt.} \\ \textit{Std.} \end{tabular}} \\ \hline
 &  &  & 0 & 19.88 & 6.51 & \cellcolor[HTML]{FFCCC9}\textbf{22.61\%} & 0.23\% & \textbf{-} & {\color[HTML]{CCCCCC} -} & 6.08 & 1.79 \\
 &  &  & 1 & \cellcolor[HTML]{DAE8FC}\textbf{7.65} & \textbf{6.52} & 0.00\% & \textbf{0.10\%} & 6.08 & \textbf{1.96} & 6.09 & 1.85 \\
 &  &  & 2 & 8.20 & 6.68 & 0.00\% & 0.01\% & 5.95 & 1.89 & 6.19 & 1.84 \\
 &  & \multirow{-4}{*}{\medium} & 3 & 9.01 & 7.20 & 0.00\% & 0.04\% & 6.01 & 1.91 & 6.12 & 1.86 \\
 &  &  & 0 & 32.08 & 4.99 & \cellcolor[HTML]{FFCCC9}\textbf{16.00\%} & 0.07\% & \textbf{-} & - & 6.11 & 1.70 \\
 &  &  & 1 & \cellcolor[HTML]{DAE8FC}\textbf{11.53} & 9.27 & 0.00\% & 0.19\% & 6.13 & 2.08 & 6.13 & 2.08 \\
 &  &  & 2 & 14.19 & 10.06 & 0.00\% & 0.02\% & 6.25 & 1.93 & 6.34 & 1.90 \\
 & \multirow{-8}{*}{$C_1$} & \multirow{-4}{*}{\short} & 3 & 14.92 & 10.65 & 0.00\% & 0.01\% & 6.12 & 1.88 & 6.39 & 1.89 \\
 &  &  & 0 & 31.17 & 11.06 & \cellcolor[HTML]{DAE8FC}0.00\% & 0.15\% & \textbf{-} & - & 6.15 & 1.70 \\
 &  &  & 1 & 15.98 & 12.22 & 0.00\% & 0.06\% & 6.29 & 1.89 & 6.12 & 1.93 \\
 &  &  & 2 & 15.23 & 11.93 & 0.00\% & 0.01\% & 6.22 & 1.84 & 6.33 & 1.78 \\
\multirow{-12}{*}{\textit{\begin{tabular}[c]{@{}c@{}}CodeLlama2\\ 7B\end{tabular}}} & \multirow{-4}{*}{$C_2$} & \multirow{-4}{*}{\short} & 3 & \cellcolor[HTML]{DAE8FC}\textbf{14.05} & 11.39 & 0.00\% & 0.01\% & 6.07 & 1.93 & 6.17 & 1.77 \\ \hline
 &  &  & 0 & 28.73 & 1.01 & \cellcolor[HTML]{FFCCC9}\textbf{41.69\%} & 1.10\% & \textbf{-} & - & 6.22 & 1.85 \\
 &  &  & 1 & 28.85 & 5.22 & 11.11\% & 1.21\% & 6.14 & 2.02 & 6.09 & 1.87 \\
 &  &  & 2 & \cellcolor[HTML]{DAE8FC}\textbf{26.26} & 5.26 & 10.10\% & 2.40\% & 6.02 & 1.90 & 5.95 & 1.82 \\
 &  & \multirow{-4}{*}{\medium} & 3 & 27.23 & 5.03 & 11.71\% & \cellcolor[HTML]{FFCCC9}2.58\% & 5.94 & 1.97 & 6.10 & 1.82 \\
 &  &  & 0 & \cellcolor[HTML]{DAE8FC}\textbf{25.97} & 1.98 & \cellcolor[HTML]{FFCCC9}\textbf{80.33\%} & 0.92\% & \textbf{-} & - & 6.25 & 1.80 \\
 &  &  & 1 & 31.98 & 4.49 & 15.72\% & 0.60\% & 6.23 & 1.85 & 6.11 & 1.87 \\
 &  &  & 2 & 35.31 & 2.50 & 27.27\% & 0.10\% & 6.22 & 1.87 & 6.33 & 1.68 \\
 & \multirow{-8}{*}{$C_1$} & \multirow{-4}{*}{\short} & 3 & 31.48 & 3.17 & 35.25\% & \cellcolor[HTML]{FFCCC9}1.62\% & 6.11 & 1.88 & 6.38 & 1.75 \\
 &  &  & 0 & 37.50 & \textbf{0.71} & \cellcolor[HTML]{DAE8FC}\textbf{44.30\%} & 0.44\% & - & - & 6.20 & 1.81 \\
 &  &  & 1 & \cellcolor[HTML]{DAE8FC}\textbf{21.72} & 3.67 & 47.80\% & 0.90\% & 6.32 & 1.77 & 6.20 & 1.87 \\
 &  &  & 2 & 27.60 & 1.80 & 60.61\% & 1.13\% & 6.26 & 2.02 & 6.08 & 1.78 \\
\multirow{-12}{*}{\textit{\begin{tabular}[c]{@{}c@{}}Mistral\\ 7B\end{tabular}}} & \multirow{-4}{*}{$C_2$} & \multirow{-4}{*}{\short} & 3 & 30.89 & 1.14 & 59.93\% & \cellcolor[HTML]{FFCCC9}1.55\% & 6.20 & 1.85 & 6.22 & 1.97 \\ \hline
 &  &  & 0 & 10.55 & 9.98 & 0.00\% & \cellcolor[HTML]{FFCCC9}5.72\% & - & - & 6.17 & 1.68 \\
 &  &  & 1 & 10.15 & 9.32 & 0.00\% & 0.73\% & 6.06 & 1.77 & 6.12 & 1.84 \\
 &  &  & 2 & \cellcolor[HTML]{DAE8FC}\textbf{9.27} & \textbf{8.66} & 0.00\% & \textbf{0.01\%} & 6.07 & 1.87 & 5.96 & 1.89 \\
 &  & \multirow{-4}{*}{\medium} & 3 & 10.50 & 9.22 & 0.00\% & 0.00\% & 5.92 & 1.94 & 6.20 & 1.86 \\
 &  &  & 0 & 16.19 & 13.71 & 0.00\% & \cellcolor[HTML]{FFCCC9}8.44\% & - & - & 6.36 & 1.82 \\
 &  &  & 1 & 15.02 & 12.43 & 0.00\% & 0.00\% & 6.21 & 1.98 & 6.26 & 1.84 \\
 &  &  & 2 & \cellcolor[HTML]{DAE8FC}\textbf{14.36} & 11.72 & 0.00\% & 0.00\% & 6.24 & 1.81 & 6.03 & 1.78 \\
\multirow{-8}{*}{\textit{\begin{tabular}[c]{@{}c@{}}CodeLlama2\\ 13B\end{tabular}}} & \multirow{-8}{*}{$C_1$} & \multirow{-4}{*}{\short} & 3 & 15.15 & 12.83 & 0.00\% & 0.00\% & 6.31 & 1.85 & 6.22 & 1.83 \\ \hline
\end{tabular}
}
{\\ \footnotesize *Best \metricMSE in blue-bold. Worst non-feasible and out-of-range proportions in red.}
\end{table}

\rev{To address \ref{rq:best_config}, we also tested our models at $T=0.9$ to capture the largest model outcome range. Table~\ref{tab:mse-03} presents the \metricMSE for \codellama and \mistral models. We observe a higher \metricNonFeasible proportion for $C_1$ than $C_2$ for both models \codellamaS and \mistral with zero shots; for example, \codellamaS has a \metricNonFeasible } value of $16\%$ and $80.33\%$ for \mistral with $C_1$ and $0\%$ with $C_2$. 

In comparison, we observe no big difference in the accuracy \metricMSE between using $C_1$ and $C_2$.\codellamaS with 3-shots and $C1$ has an \metricMSE of $14.92$ and a Std. of $10.65$ while the same configuration as $C_2$ has a \metricMSE of $14.05$ and a Std. of $11.39$.

\mistral obtains the best \metricMSE of $21.72$ with $C_2$; however, the proportion of non-feasible ($47.80\%$) is high. We notice that the role and task description of $C_2$ improves the non-feasible proportion to the zero-shot configuration by observing $80.33\%$ with $C_1$ and $44.3\%$ with $C_1$ using \mistral. Similarly, \codellamaS has a \metricNonFeasible proportion of $16\%$ zero shot and $C_1$, and $0\%$ with zero-shot and $C_2$.
 


\rev{\textbf{Discussion.} Our findings reveals positive impact of using $C_3$ compared to $C_1$ and slight improvement of using $C_3$ compared with $C_2$ when using \codellama model with $0.3$ temperature. Comparing the same context samples, we observe an impact on the number of shots; \textit{Accuracy jumps} observed from zero-shot to 1-shot indicate that LLMs obtain more effective guidance by providing examples rather than explaining the role (\ie the CVSS is a severity value that ranges between 0 to 10) and describing the task (\ie your roles is to examine the code and predict a score). We observe no significant difference between $C_1$ and $C_2$ at $0.9$ temperature. Instead, $C_2$ positively impacts the prediction by reducing the \metricNonFeasible proportion in the predicted score. However, the model potentially produces inaccurate predictions outside the correct score range. }
\begin{figure*}[t]
\centering
\begin{subfigure}[t]{0.32\textwidth}
    \centering
      \includegraphics[width=\linewidth]{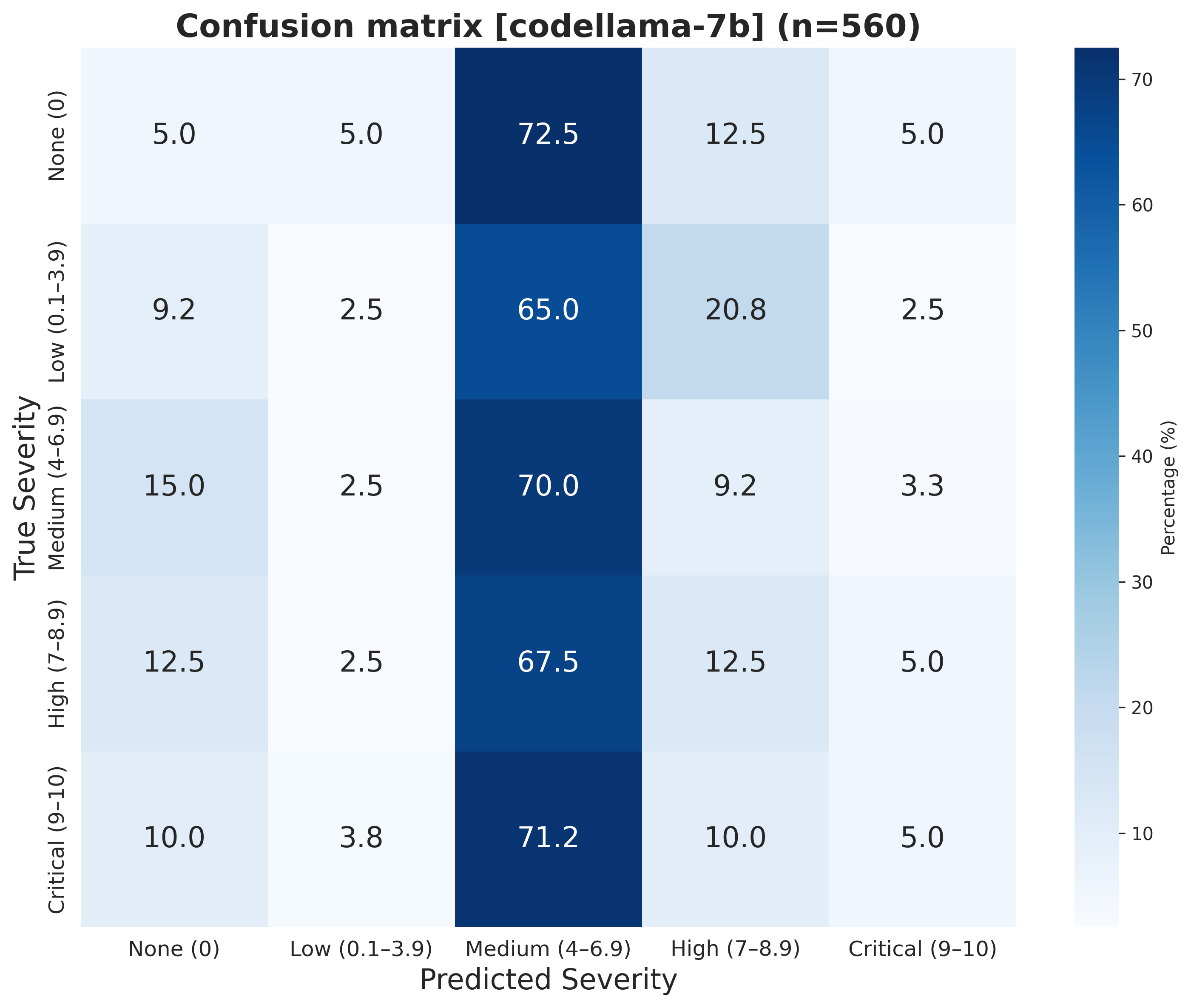}

\end{subfigure}
\hfill
\begin{subfigure}[t]{0.32\textwidth}
    \centering
    \includegraphics[width=\linewidth]{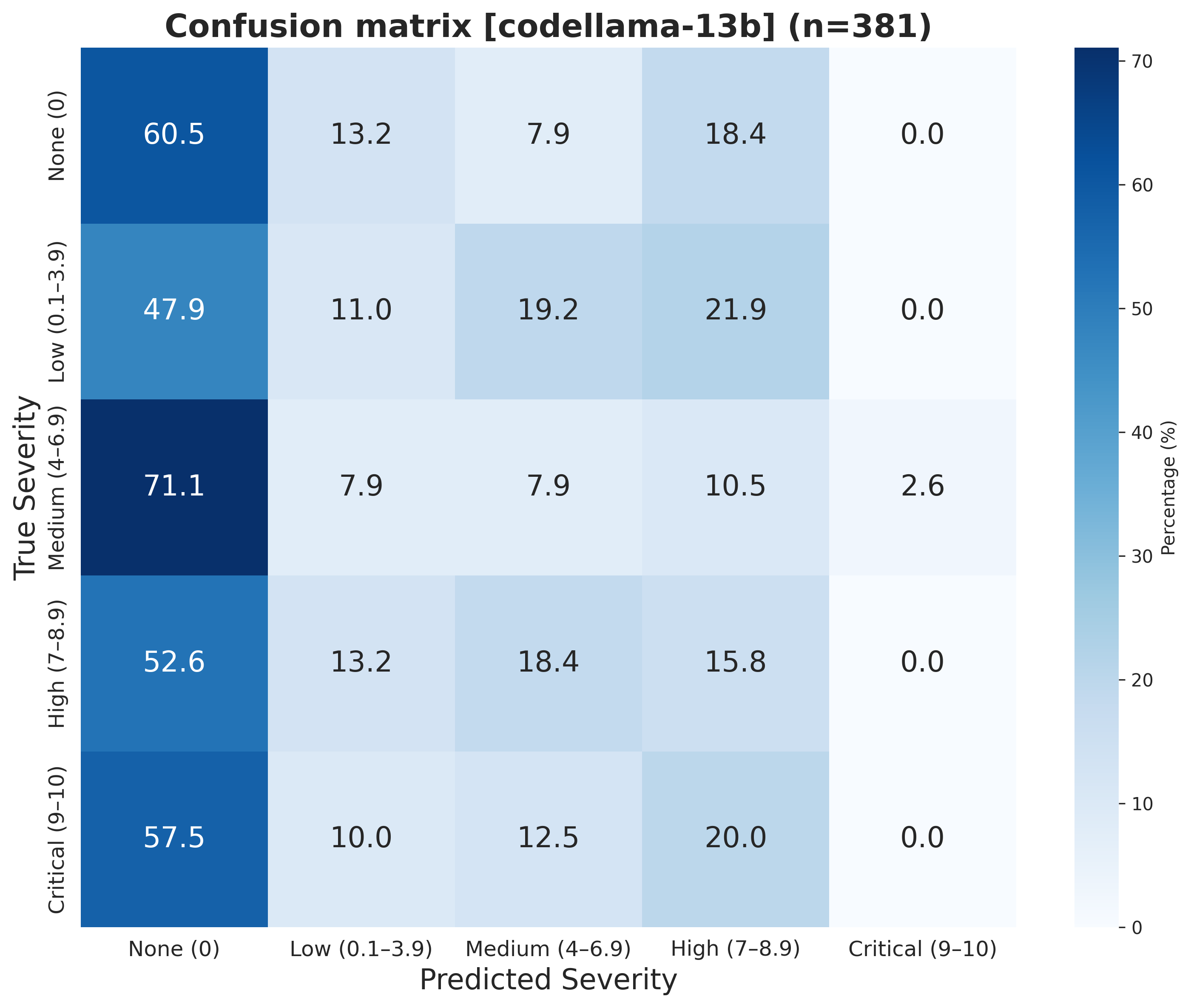}

\end{subfigure}
\hfill
\begin{subfigure}[t]{0.32\textwidth}
    \centering
    \includegraphics[width=\linewidth]{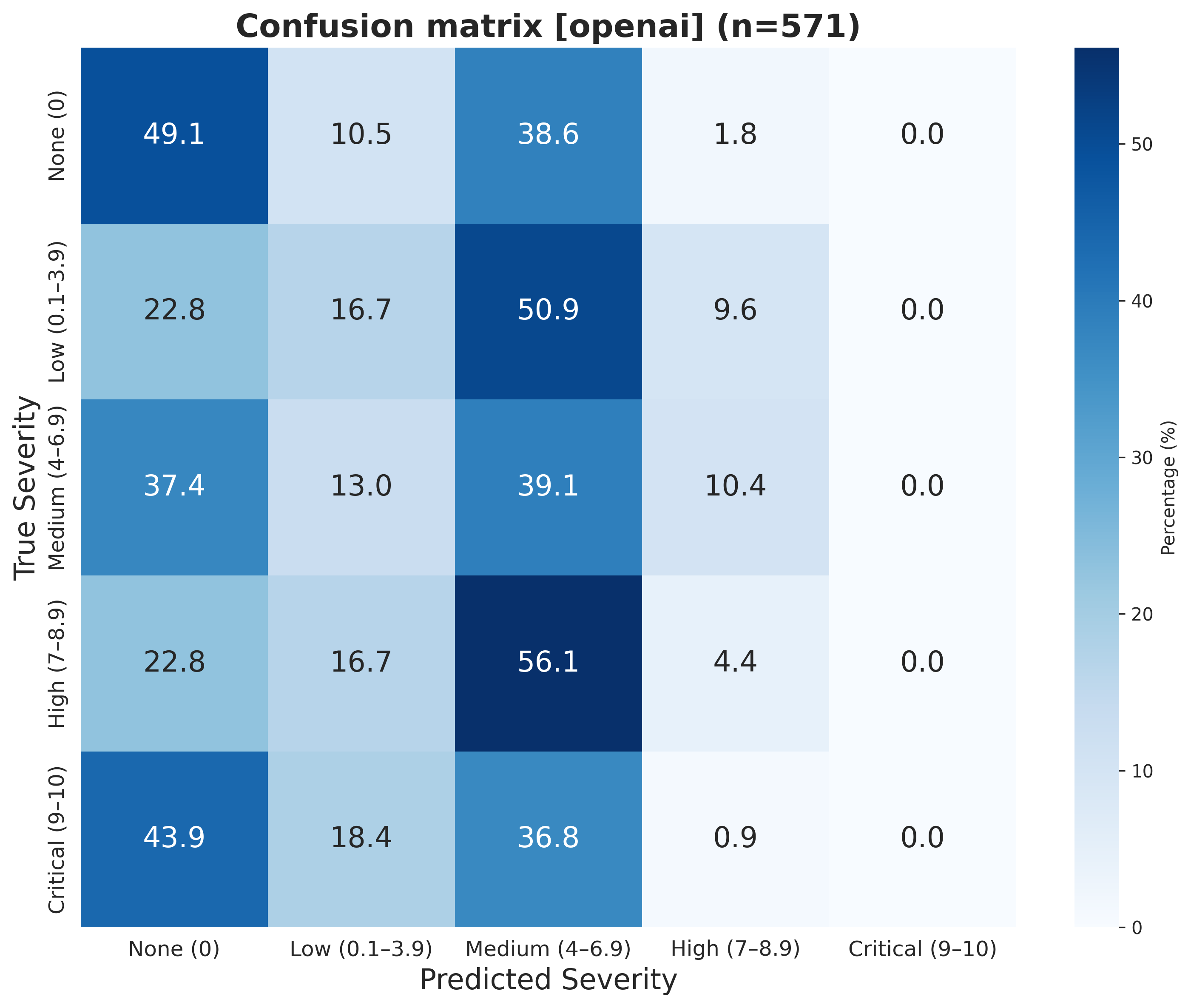}
   
\end{subfigure}
\caption{Confusion matrix for stratified CVSS classification using \codellamaS, \codellamaL and \gptf with $n=600$ datapoints.}
    \label{fig:confusion_matrix}
\end{figure*}

\section{RQ$_3$ Open-source LLMs Accuracy}

\rev{The \metricMSE alone does not provide sufficient reliability and therefore requires a complementary measure. To address this, we use \metricNonFeasible, which captures whether at least one valid answer emerges within 30 attempts. For example, \mistral with zero-shot prompting and \short sequence length achieves a \metricMSE of $25.97$, lower than that obtained with \medium sequences under the same prompting condition. However, this improvement coincides with a substantially higher rate of feasible answers: \metricNonFeasible increases from $41.69\%$ for \medium sequences to $80.33\%$ for \short sequences. We run experiments on each model \codellamaS and \codellamaL, \mistral with $C_1$ varying the prompt size and the number of shots. \tabref{tab:mse-03} and \tabref{tab:socre_evaluation} illustrate the regression and standard deviation metrics computed for the experiment. We observe that \codellama models outperform \mistral in all sequence sizes and have lower \metricNonFeasible rates. }

\rev{In contrast, the \codellama models have a lower number of \metricNonFeasible responses. We observe that \codellama on zero-shot does not have enough context in the score range. Therefore, the \metricOutofRange percentage increases to $8.44\%$ for a sequence length of 100. Note that the best setup is the \codellamaS model size with just one-shot and sequence size of 300, obtaining an \textit{Avg.} \metricMSE of $7.65$ and a \textit{Std.} of $6.52$ with a $0.10\%$ \metricOutofRange values. Similar behavior is observed with \codellamaS with a sequence size of 300 and two shots observing an \textit{Avg.} \metricMSE of $9.27$ and a \textit{Std.} of $8.66$. }

\rev{All models, except \codellamaL, produce a significant number of \metricNonFeasible responses with a temperature of $0.9$. Among them, \mistral in the zero-shot setting performs the worst, with 80.33\% of its responses being \metricNonFeasible. This result indicates that the model requires more context and a clearer description of the expected output to generate acceptable outcomes. In contrast, \codellamaL shows a consistent ability to predict feasible CVSS values without the need for additional context. A similar pattern is observed with \codellamaS; however, its performance depends on the specific example provided. For example, when evaluated with $C_1$ in the zero-shot setting, \codellamaS occasionally fails to generate a valid value, highlighting its sensitivity to the input example.}

\rev{Figure~\ref{fig:confusion_matrix} presents confusion matrices summarizing the severity classification performance of the three evaluated models (\codellamaS, \codellamaL, and \gptf) on the held‑out test set, with rows denoting true severity and columns the predicted severity.}

\rev{The \codellamaS model (left) concentrates predictions in the ``Medium'' column, leading to high correct rates for true “Medium” cases but systematic under‑prediction of “None” and “Low” severities. This tendency to default to “Medium” results in fewer extreme false positives, but at the cost of reduced discrimination among lower-severity classes. For \codellamaL (center), has a tendency to predict scores to the "None" category. The matrix also reveals a notable misclassification in the most extreme “Critical” category, suggesting conservative behavior when assigning the highest severity.}

\rev{The \gptf model (right) shows comparatively stronger concentrations for the “None” and “Medium” classes, indicating a slightly better alignment between "Medium" and predicted labels in these regions. At the same time, the model fails to predict "Critical" labels, producing "None" or "Medium" classifications. Similarly to \codellamaL, the model almost never predicts “Critical,” suggesting a cautious stance toward the highest severity assignments.}

\textbf{Discussion.} We observe that \codellama models outperform in predicting a score value given a prompt with vulnerable code and examples; \codellama models tend to predict a value and then explain why the value is valid. \codellama models increase the verbosity of the generated text as the size of the model increases. Consequently, the \textit{Std.} for 13B is larger than the ones for 7B. For example, \codellamaS, with a three-shot sequence size of 300, has a variation of $10.65$, while the same experiment for 13B has a variation of $12.83$. In contrast, \mistral is prone to summarize code execution before predicting the severity score.  

We observe a small difference between the model sizes' performance. In fact, the small \codellamaS has better accuracy in completing the scoring prediction tasks with a few \metricOutofRange values. We highlight that the \metricMSE values are comparable between \llms and experiments, since we sampled from the same \bigvul data set and observe very close values between experiments for \metricExMean, \metricExStd, \metricGTMean, and \metricGTStd 
 \rev{The matrix structure suggests that \codellamaS is less calibrated across the full severity range, effectively blending several neighboring classes into a single dominant prediction.}

\section{Related Work}\label{sec:related_work}

\rev{Prior work on learning-based vulnerability analysis has focused on traditional machine learning or pre-LLM deep models for detection and severity estimation, for example, linear models and random forests for CVSS prediction from textual descriptions, or CNN/RNN/Transformer-based models trained on code-level datasets such as \bigvul~\cite{fan_cc_2020}. More recently, deep vulnerability-focused models such as VulDeePecker~\cite{li_vuldeepecker_2018}, VulBERTa~\cite{hanif_vulberta_2022}, and LineVul~\cite{fu_linevul_2022} have shown that specialized architectures and pretraining objectives can improve code-level vulnerability detection, but they still require task-specific training and do not directly address CVSS regression. In contrast, a growing body of work now studies how general-purpose LLMs can be adapted to security tasks, including LLM-assisted static analysis (IRIS)~\cite{li2025iris}, LLM-based data-flow reasoning (LLMDFA)~\cite{wang2024llmdfa}, CVE/CVSS classification~\cite{marchiori2025llmsclassifycvesinvestigating}, LLM-based vulnerability detection, classification, and repair~\cite{fu_chatgpt_2023,khare2025effectiveness}, and critiques of current ML benchmarking practices for vulnerability detection~\cite{risse2025scorewrongexambenchmarking}. These studies highlight both the promise and the limitations of LLMs for finding bugs and reasoning about program behavior, yet they primarily treat vulnerability detection as a classification problem and often focus on cloud-hosted or fine-tuned models. Our work complements this line by targeting severity regression from vulnerable code using in-context learning with locally deployable, open-source code LLMs, and by quantifying their calibration and robustness under different prompt configurations.}

\section{Threats to Validity}
\rev{Our study operates on \bigvul’s function‑level representation of vulnerabilities, which abstracts away much of the surrounding project and operational context. As a result, our results should be interpreted as performance in these isolated snippets rather than in fully context‑rich real‑world cases.}

\rev{\textbf{Construct Validity} Our study assumes that function-level snippets from Big-Vul contain sufficient information to predict CVSS severity. However, recent work shows that method-level vulnerability labels in Big-Vul can be noisy and may not capture the full vulnerability context, which can bias models towards learning superficial correlations rather than true vulnerability semantics~\cite{risse_topscore,benchmark_vuln_ml}. We partially mitigate this by filtering for longer snippets with descriptions, but our results should still be interpreted as performance on Big-Vul's labeling rather than on arbitrary real-world code. In addition, we predict only the aggregate CVSS score, not its individual submetrics, which limits interpretability and actionability for practitioners. }

\rev{\noindent \textbf{Internal Validity}
Our results depend on specific choices for prompt format, number of shots, sequence size, and decoding temperature. We explore several configurations and, in the revised version, add a temperature calibration experiment comparing $T=0.001$ and $T=0.3$ with context $C_3$, adopting $T=0.3$ as a low-temperature setting that balances error and output diversity. However, better configurations may exist, and our reported numbers should be seen as indicative rather than optimal. We also clarify that high-temperature, multi-sample runs are used only for analyzing variability, while realistic usage should rely on a single, low-temperature prediction per query.}

\rev{\noindent \textbf{External Validity} We evaluate only C/C++ code from Big-Vul and open-source models in the 7B--13B range (CodeLlama2, Mistral). Our findings may not be generalized to other languages, vulnerability types, larger or more recent models (\eg Llama~3, Qwen~2.5), or different industrial environments. Proprietary datasets are used solely for exploratory distribution comparison, so we cannot claim end-to-end effectiveness on those systems. Moreover, recent work warns that standard vulnerability benchmarks can overestimate real-world performance due to benchmark design and spurious correlations~\cite{risse_topscore,benchmark_vuln_ml}.}

\rev{\noindent \textbf{Data Contamination and Benchmarking Bias} Big-Vul is built from public CVEs and CVSS scores, which likely appear in the pretraining data of our open-source models. This creates a risk of data contamination, where some examples may be partially memorized, increasing performance estimates~\cite{data_contam_overview,data_contam_llm}. Because we cannot audit the pretraining corpora, we treat our numbers as an upper bound under possible contamination rather than a clean measure of out-of-distribution generalization, in line with recent discussions on contamination in LLM benchmarks~\cite{data_contam_llm}.}
\section{Conclusions and Future Work}

We applied our approach as our first experience in an industry experience in a real environment. The motivation relies on creating a private and secure environment for using proprietary data and software code. \rev{Using \noncloud small and locally deployed models. This \noncloud} configures the main alternative to private options such as GitHub Copilot or OpenAI models. Our findings show a good impact on the use of \noncloud models in achieving this. 

\rev{The use of an in-context learning approach for local models further enhances this capability by enabling them to adapt dynamically to each company's specific security policies and requirements. Our approach demonstrates that acceptable performance can be achieved even with small models such as \codellamaS, striking a balance between computational efficiency and effective policy enforcement.}


\textbf{Limitations.} \rev{We predict only the aggregate CVSS v3.1 base score from C/C++ snippets, not individual submetrics or richer contextual factors (\eg infrastructure, asset criticality, social signals). Our experiments are constrained to \bigvul and to 7B–13B open-source models (\codellamaS/\codellamaL, \mistral) that fit single-GPU, \textit{on-premise deployments}, so larger recent open or proprietary models (\eg Llama 3, Qwen 2.5 32B, DeepSeek R1) remain outside our scope, and we do not offer a comprehensive LLM benchmark.}

\noindent \textbf{L$_1$: Datasets include obsolete or poor quality vulnerability description.} We observe a similar distribution and behavior between the \cisco and open datasets. However, the data can sometimes be outdated, including information from previous years. Since vulnerabilities and attacks on software systems evolve more rapidly than their corresponding reports, the available data points for training and testing models are often insufficient to meet the demands of modern cybersecurity challenges.

\noindent \textbf{L$_2$: \llms do not improve accuracy after 3--shots.} We observe \rev{a noticeable improvement in} \metricMSE predictions and feasibility at the one-shot setting. However, beyond two shots, \llms do not \rev{appear to gain sufficient additional context to meaningfully improve accuracy. Instead, higher accuracy is achieved with $C_3$ when the desired outcome format is explicitly specified. We did not evaluate additional shots due to the computational cost and the marginal improvement observed between the two-shot and three-shot settings.}

\noindent \textbf{L$_3$: \llms can assist security analysts in identifying vulnerable code before deployment.} Small and medium-sized companies can benefit from using \noncloud models while keeping their data custody and saving costs due to local execution fitting on medium GPU requirements. However, predicting severity requires larger context variables {(\eg third-party libraries, vulnerability social media information, similar vulnerabilities information)} and in-context configurations to the environmental customer variable
{(\eg infrastructure, data protection policies, data profile)}.  
Both \codellama and \mistral \llms tend to generate explanations about the code function. The explanation can be used to calibrate the model and also to track the vulnerability report.
\\

\rev{\textbf{Future work.} An immediate extension is to move beyond a single-target regression setup and jointly model CVSS alongside related risk signals such as Kenna Risk and EPSS, enabling multi-target inference that better reflects real-world prioritization pipelines. Decomposing the task to predict CVSS submetrics (\eg attack vector, privileges required, user interaction, impact metrics) in addition to the aggregate score could also yield more interpretable and actionable guidance for analysts. On the modeling side, evaluate stronger recent open-weight models and selected proprietary systems under realistic deployment constraints—for example, carefully comparing 7B–13B on-premise models against larger (\eg 32B–70B).}





%



\bibliographystyle{abbrv}
\bibliography{utils/main}

%


\end{document}